\documentclass{sig-alternate}

\usepackage{color}
\usepackage{booktabs}
\usepackage{dcolumn}
\usepackage{caption}

\def\shownotes{0}  
\def\full{1} 

\usepackage{xcolor}
\usepackage{marginnote}
\definecolor{myred}{cmyk}{0,1,.65,.34}

\usepackage{amsmath}
\usepackage{subfigure}

\usepackage{url}
\usepackage{caption, xspace}
\DeclareCaptionType{copyrightbox}
\usepackage{indentfirst}
\usepackage{clrscode}
\usepackage{array}
\usepackage{booktabs}
\usepackage{enumitem}
\setlist{nolistsep}
\usepackage{comment}

\newcommand{\hcald}{\hat{\mathcal D}}
\newcommand{\hcale}{\hat{\mathcal E}}

\newcommand{\var}{\mathrm{Var}}
\newcommand{\calb}{\mathcal{B}}

\newcommand{\calt}{\mathcal{T}}
\newcommand{\cald}{\mathcal{D}}
\newcommand{\supp}{\mathrm{Supp}}

\setitemize{noitemsep,topsep=0pt,parsep=0pt,partopsep=0pt}

\newtheorem{theorem}{Theorem}[section]

\newtheorem{corollary}[theorem]{Corollary}

\providecommand{\ie}{\emph{i.e.,} }
\providecommand{\eg}{\emph{e.g.,} }

\providecommand{\etc}{\emph{etc.}}      
\providecommand{\mypara}[1]{\smallskip\noindent\emph{#1} }
\providecommand{\myparab}[1]{\smallskip\noindent\textbf{#1} }

\ifnum\shownotes=1
\newcommand{\authnote}[2]{{ $\ll$\textsf{\footnotesize #1 notes: #2}$\gg$}}
\else
\newcommand{\authnote}[2]{}
\fi
\newcommand{\Znote}[1]{{\authnote{Zhenming}{#1}}}

\makeatletter
\def\@copyrightspace{\relax}
\makeatother

\begin{document}
\title{Learning about social learning in MOOCs: 
\\ From statistical analysis to generative model}
\ifnum \full=0
\numberofauthors{1} 
\fi
%
\ifnum \full=1
\numberofauthors{6} 
\fi

\author{
%
%
\ifnum \full=0
\alignauthor
Paper \# 479\\
\fi
\ifnum \full=1
\alignauthor
Christopher G. Brinton
\affaddr{Princeton University}
\affaddr{Princeton, NJ}
\email{cbrinton@princeton.edu}
\alignauthor
Mung Chiang
\affaddr{\\Princeton University}
\affaddr{Princeton, NJ}
\email{chiangm@princeton.edu}
\alignauthor
Shaili Jain
\affaddr{\\Microsoft}
\affaddr{\\Bellevue, WA}
\email{shj@microsoft.com}
\and
\alignauthor
Henry Lam
\affaddr{\\Boston University}
\affaddr{\\Boston, MA}
\email{khlam@bu.edu}
\alignauthor
Zhenming Liu
\affaddr{\\Princeton University}
\affaddr{\\Princeton, NJ}
\email{zhenming@princeton.edu}
\alignauthor
Felix Ming Fai Wong
\affaddr{\\Princeton University}
\affaddr{\\Princeton, NJ}
\email{mwthree@princeton.edu}
\fi
}

%
%
%
%
%

\maketitle

\begin{abstract}
We study user behavior in the courses offered by
a major Massive Online Open Course (MOOC) provider during the summer of 2013.  Since social learning is a key element of scalable education in MOOCs and is done via online discussion forums, our main focus is in understanding forum activities. 
Two salient features of MOOC forum activities drive our research:
1. \emph{High decline rate:} for all courses studied, the volume of discussions in the forum declines
continuously throughout the duration of the course. 2. \emph{High-volume, noisy discussions:}
at least 30\% of the courses produce new discussion threads at rates that are infeasible
for students or teaching staff to read through. Furthermore, a substantial portion
of the discussions are not directly course-related. 

We investigate factors that correlate with the decline of activity in the online discussion forums
and find effective strategies to classify threads and rank their relevance.
Specifically, we use linear regression models to analyze the time series of the count data for the forum
activities and make a number of observations, \eg the teaching staff's active participation in the discussion
increases the discussion volume but does not slow down the decline rate. We then propose a
unified generative model for the discussion threads, which allows us both to choose efficient thread classifiers and design an effective algorithm for ranking thread relevance.
Our ranking algorithm is further compared against two baseline algorithms, using human evaluation from Amazon Mechanical Turk. \\

\noindent \textbf{The authors on this paper are listed in alphabetical order. For media and press coverage, please refer to us collectively, as ``researchers from the EDGE Lab at Princeton University, together with collaborators at Boston University and Microsoft Corporation.''}
\end{abstract}

\category{\noindent G.1.6 }{Mathematics of Computing}{Least squares methods}
\category{I.2.6 }{Computing Methodologies}{Concept Learning}

\section{Introduction}\label{sec:intro}
The recent rapid development of massive online open courses (MOOCs) offered through websites such as Coursera, edX, 
and Udacity  demonstrates the potential of using the Internet to scale higher education.  Besides business models and potential impact, pedagogy is an often-debated subject as MOOCs try to make higher education available to a broader base. Low completion rates have often been cited to highlight a scale--efficacy tradeoff~\cite{Swan02,Kuh09, Bou11, VPL11}. 

Social learning is a key part of MOOC platforms. It holds the promise of scalable peer-based learning and is often the dominant channel of teacher-student interactions. As MOOCs proliferate in 2013, a natural question is: How can we better understand MOOC forum activities through the large-scale, extensive data emerging in the recent months? 
It has been observed that such forums suffered from the following major problems~\cite{kop2011challenges, QHB12}:

\begin{itemize}
\item \textbf{Sharp decline rate:} The amount of interaction in forums rapidly drops soon after a course is launched.
\item \textbf{Information overload:} As courses reach a larger audience, often the forum is flooded by discussions from many students; thus it quickly becomes infeasible for anyone to navigate the discussions to find relevant information.
\end{itemize}

\begin{figure*}
\begin{center}
\subfigure[Number of posts per day]{
\includegraphics[scale=0.25]{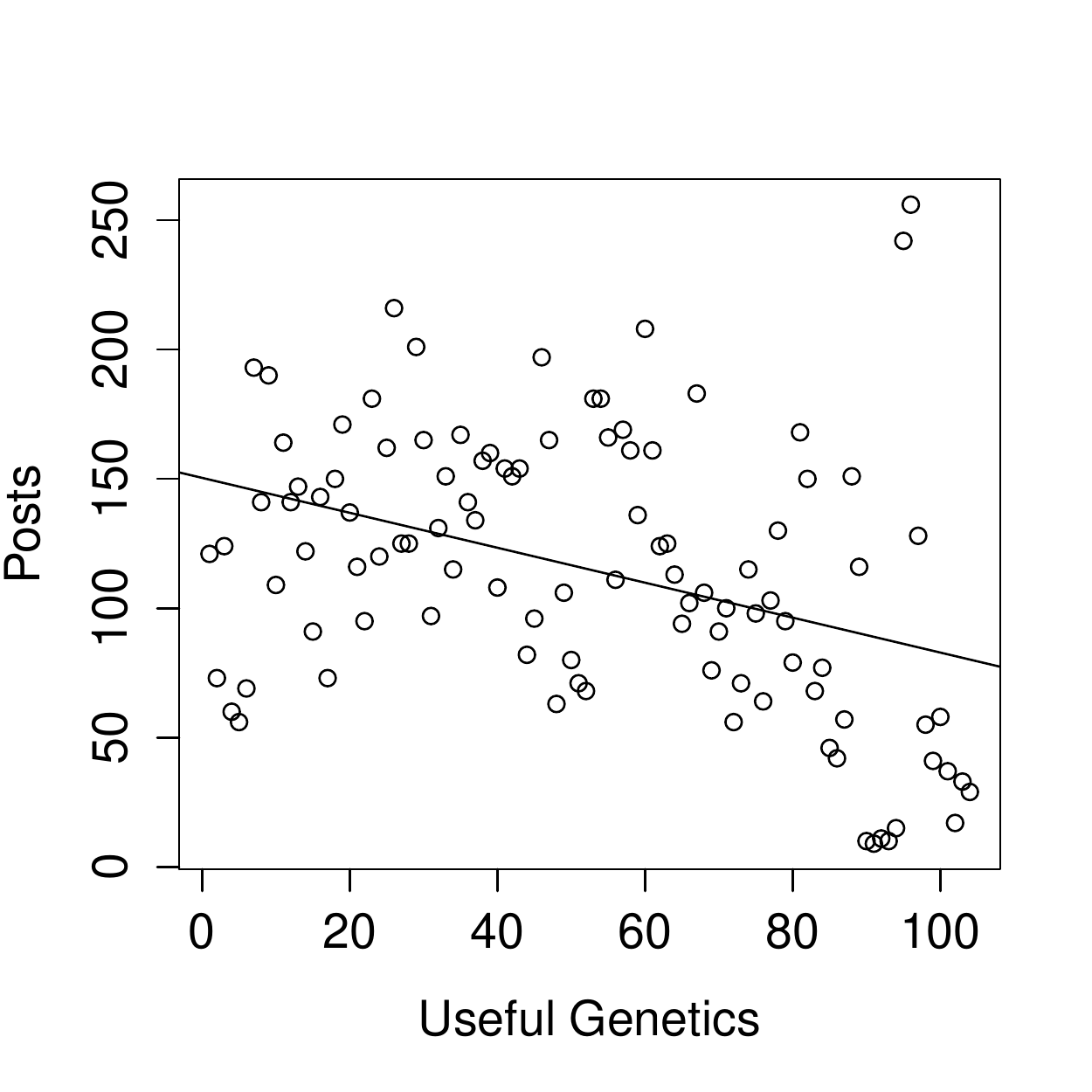}
\includegraphics[scale=0.25]{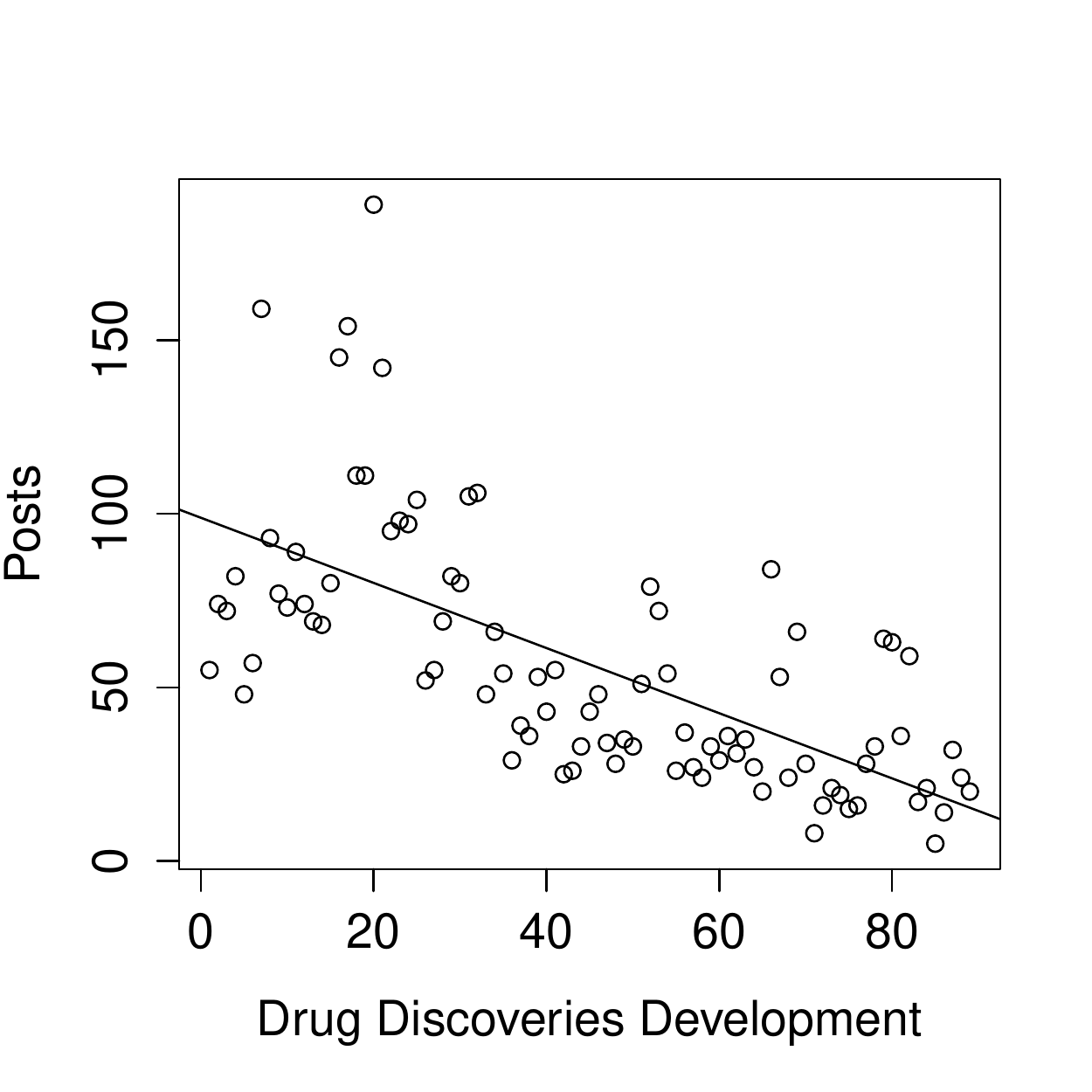}
\includegraphics[scale=0.25]{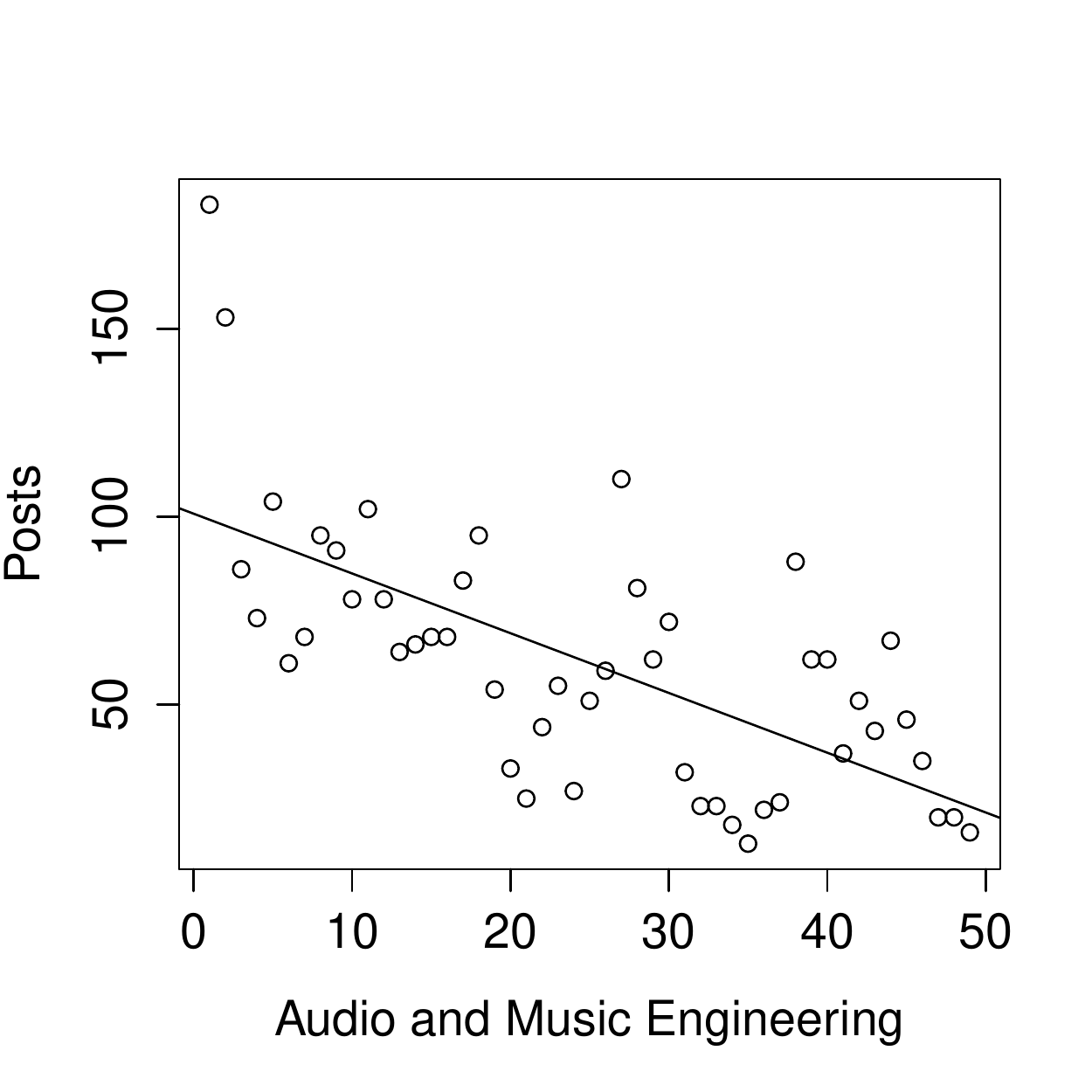}
\includegraphics[scale=0.25]{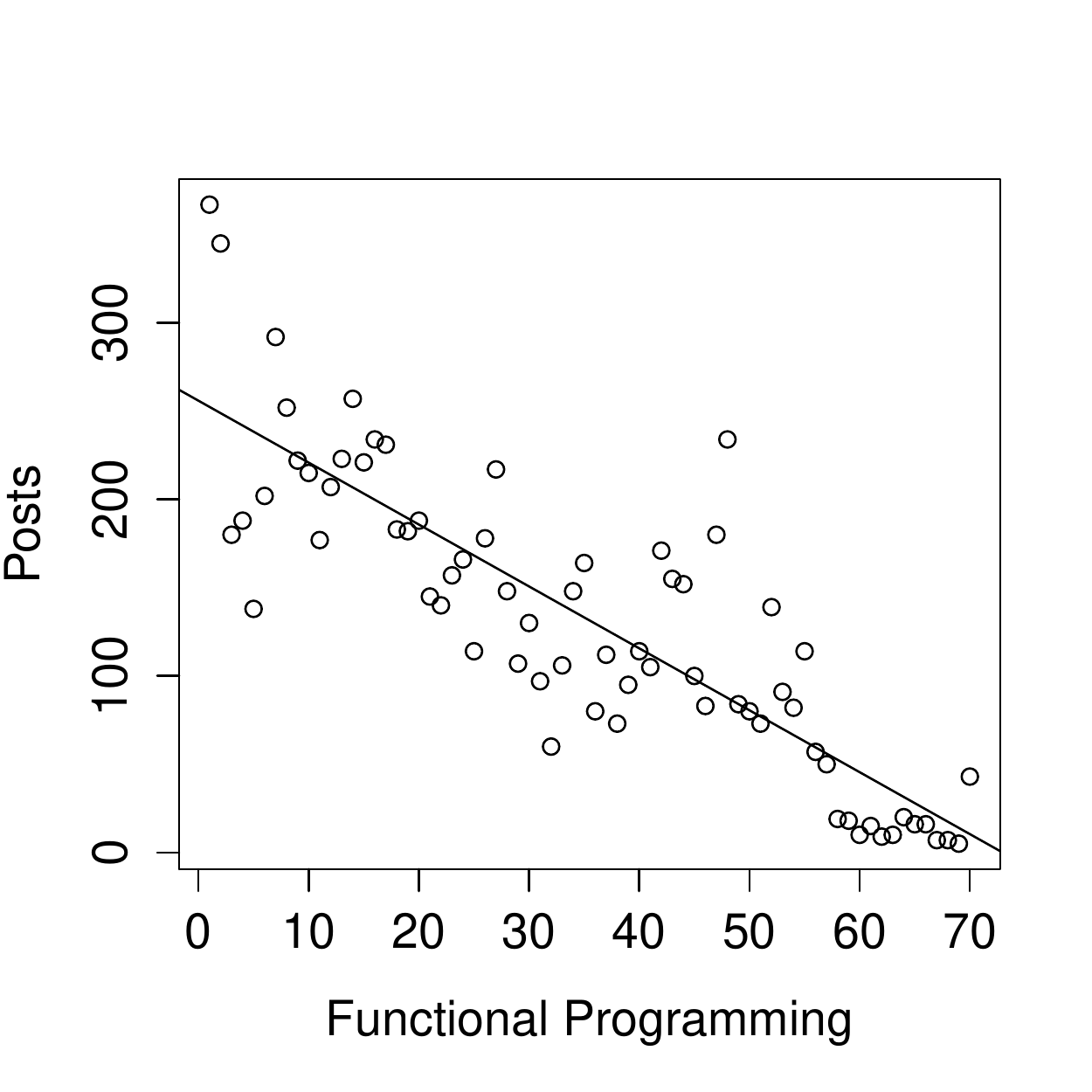}
\includegraphics[scale=0.25]{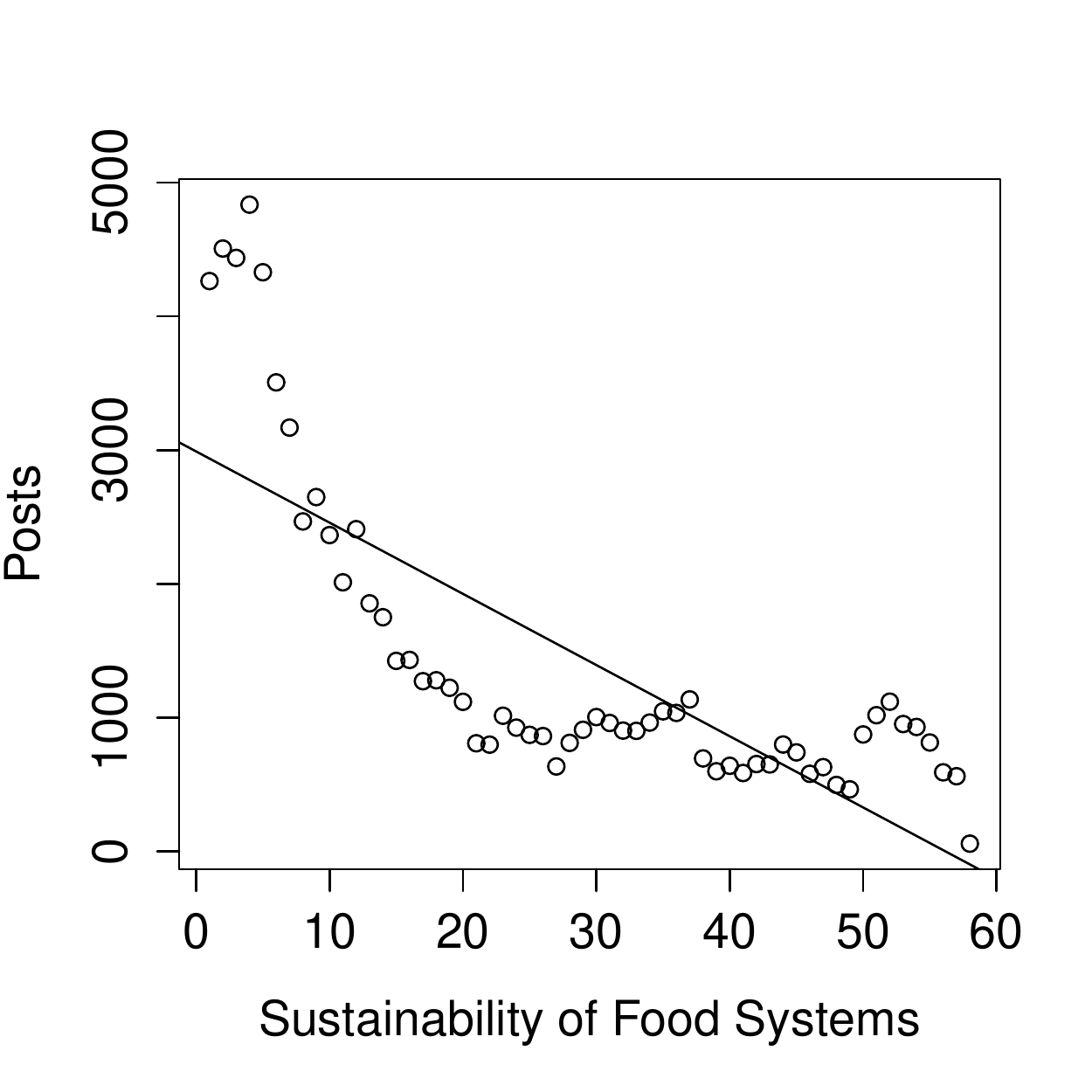}
}
\subfigure[Difference in post counts between two consecutive days]{
\vspace{-.5cm}
\includegraphics[scale=0.25]{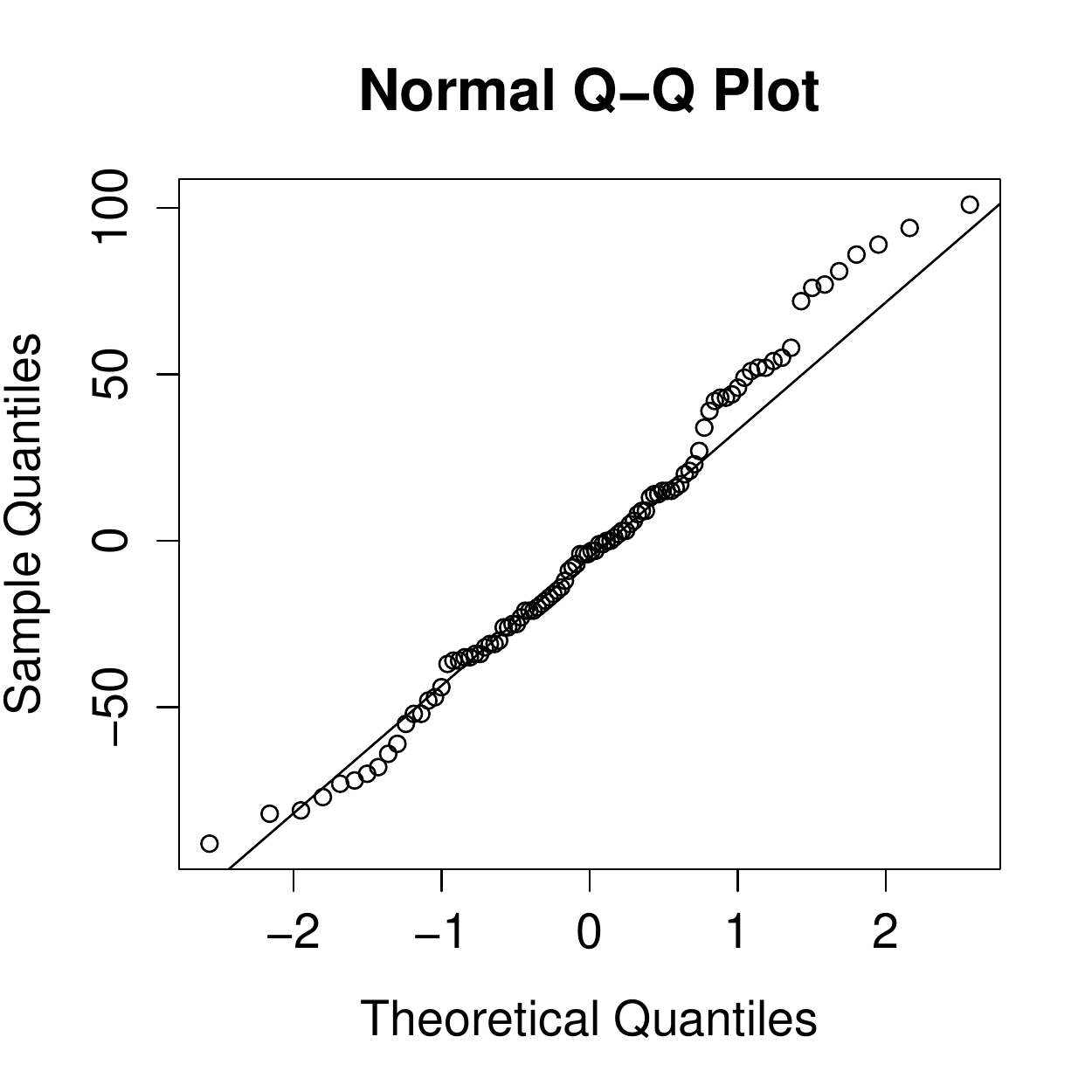}
\includegraphics[scale=0.25]{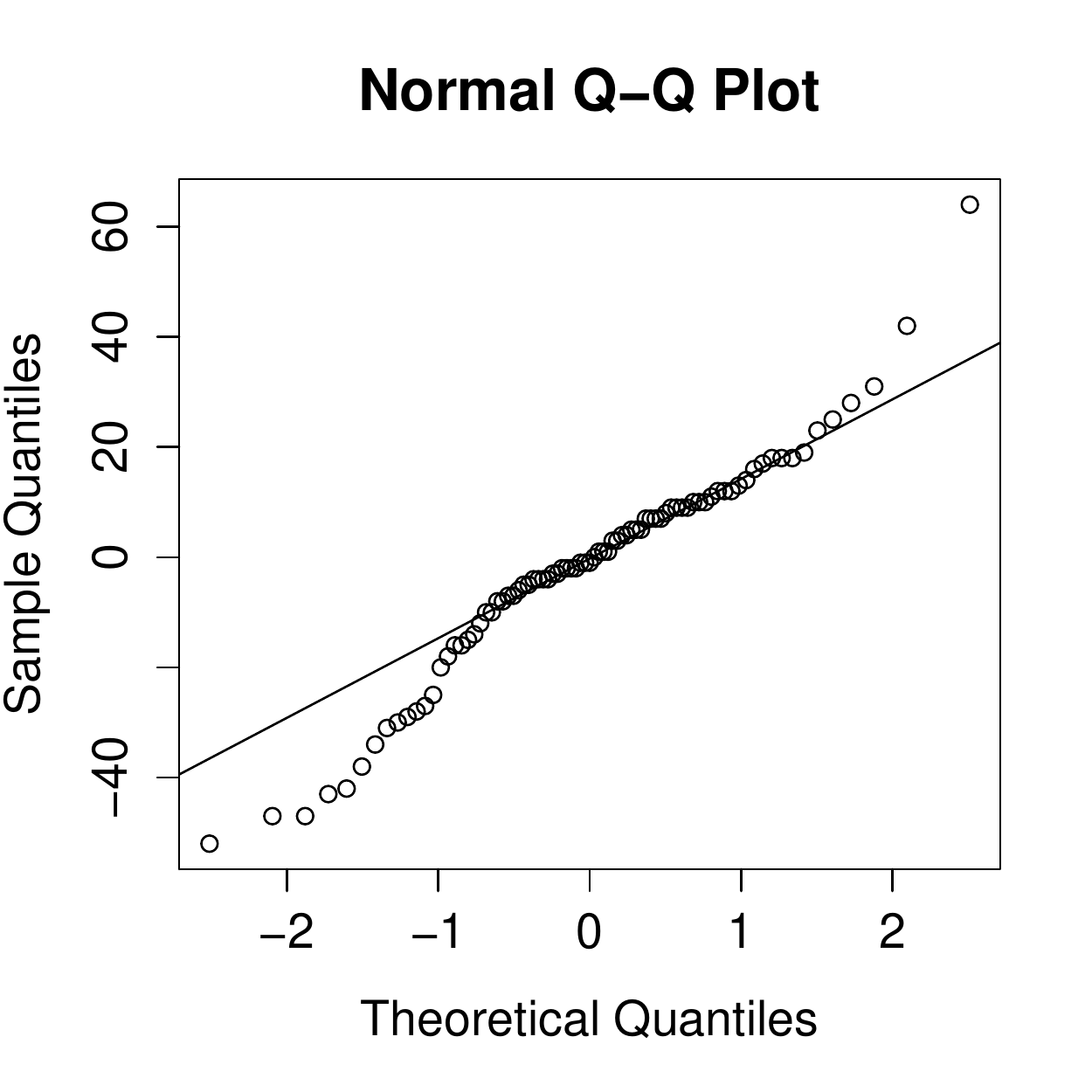}
\includegraphics[scale=0.25]{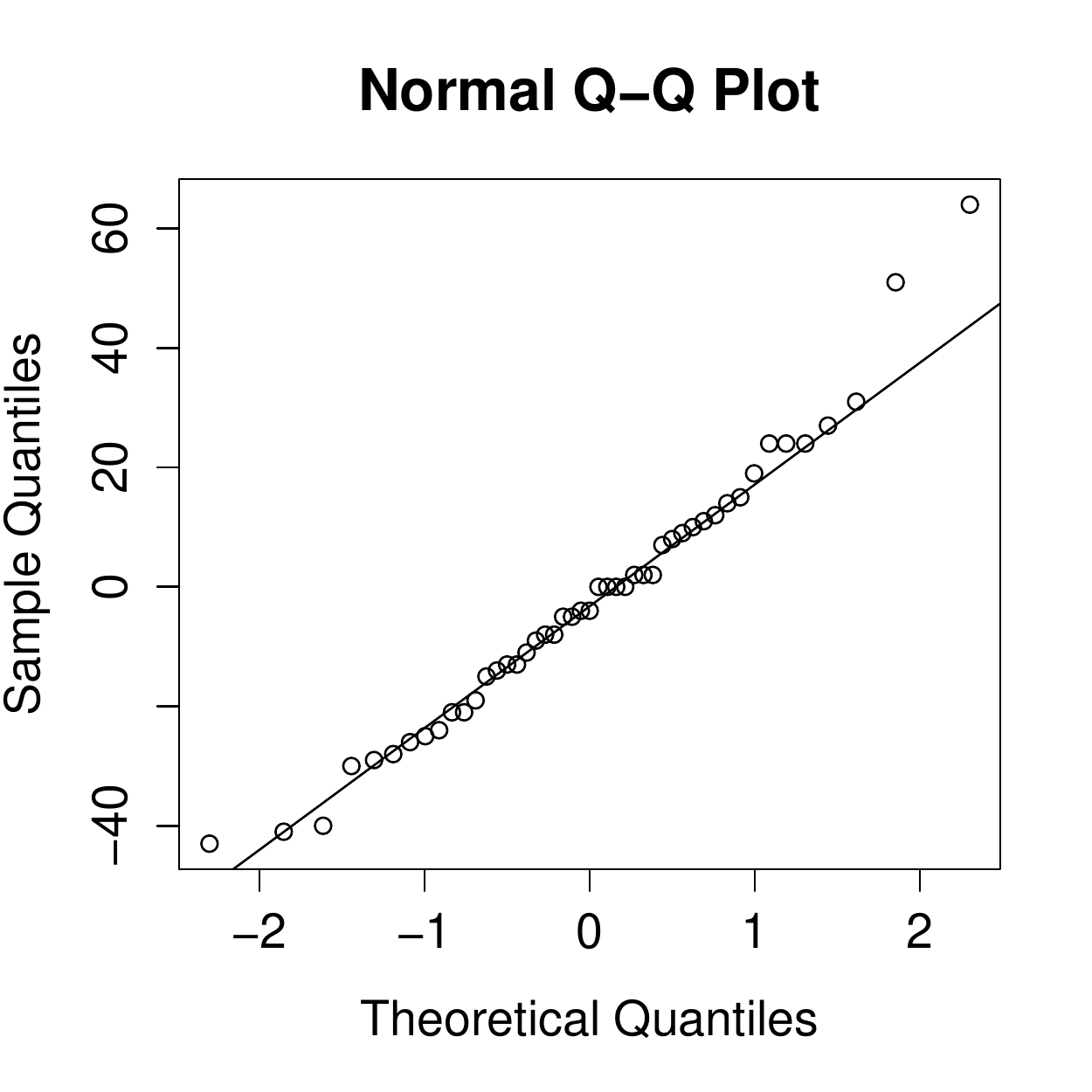}
\includegraphics[scale=0.25]{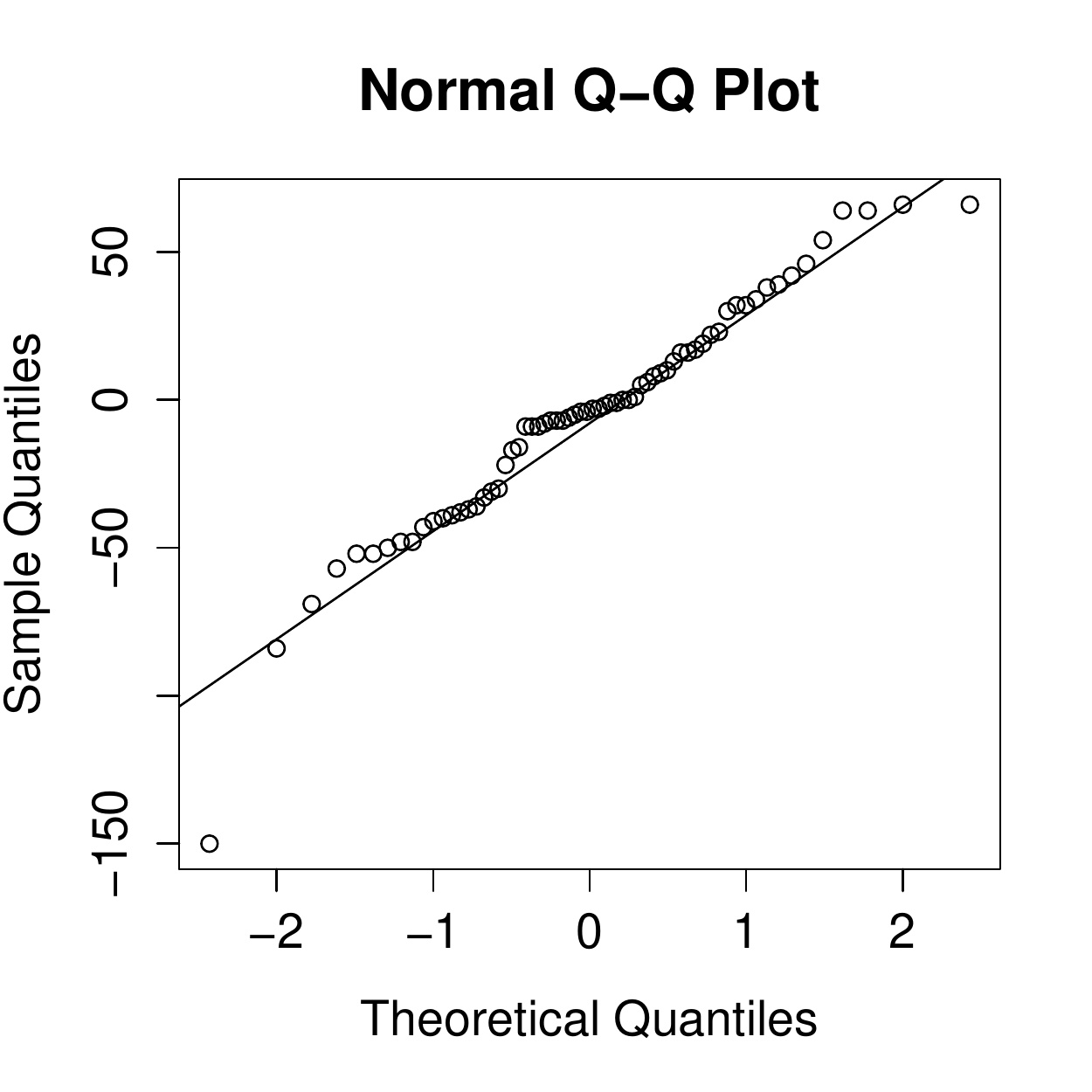}
\includegraphics[scale=0.25]{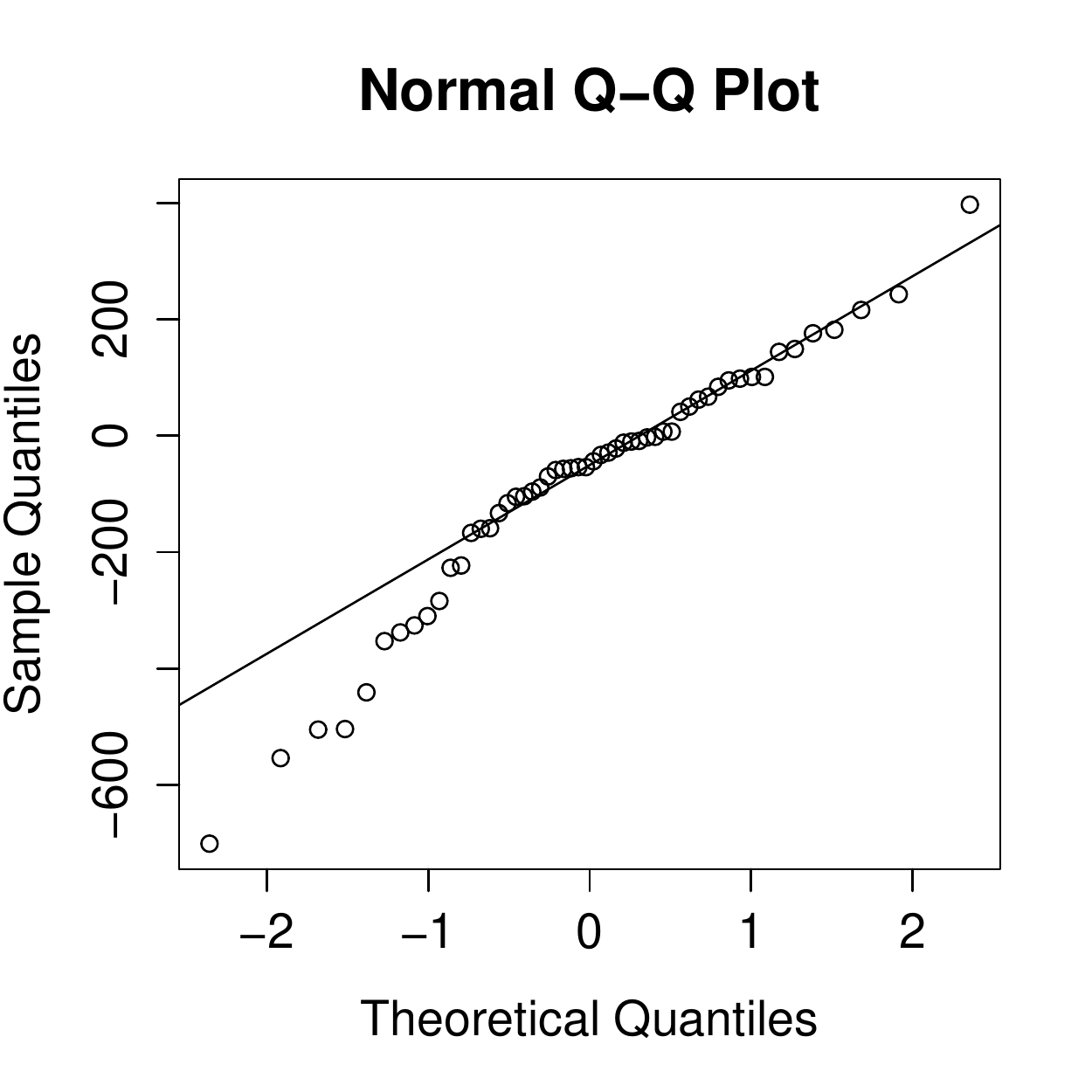}}
\end{center}
\vspace{-.5cm}
\caption{The decline of forum activities. Top: We randomly chose 5 courses out of the 73 courses we crawled and plot
the number of posts over the days. A regression line with regressor on time is added. Bottom:
 Q-Q plot on the differences in counts between two consecutive days after removing 6\% outliers (See Section~\ref{sec:regression}).}
\vspace{-.1cm}
\label{fig:declines}
\end{figure*}

\begin{figure*}
\begin{center}
\subfigure[Sampled portion of small-talk by categories]{
\includegraphics[scale=0.3]{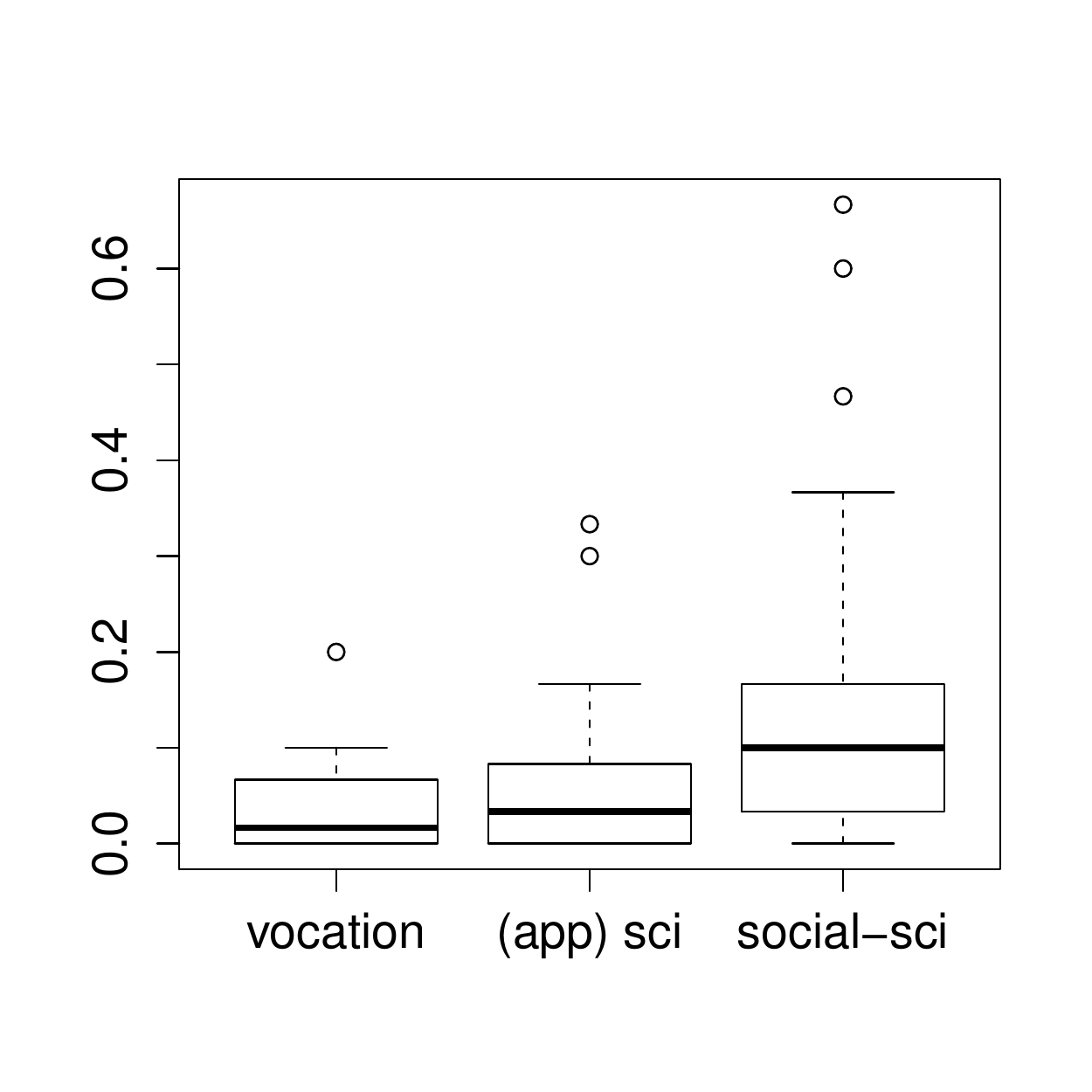}}
\subfigure[Moving averages on the portion of small-talk by categories. Left: Vocational; Mid: (Applied) Sciences; Right:  Humanities and Social Sciences .]{
\includegraphics[scale=0.34]{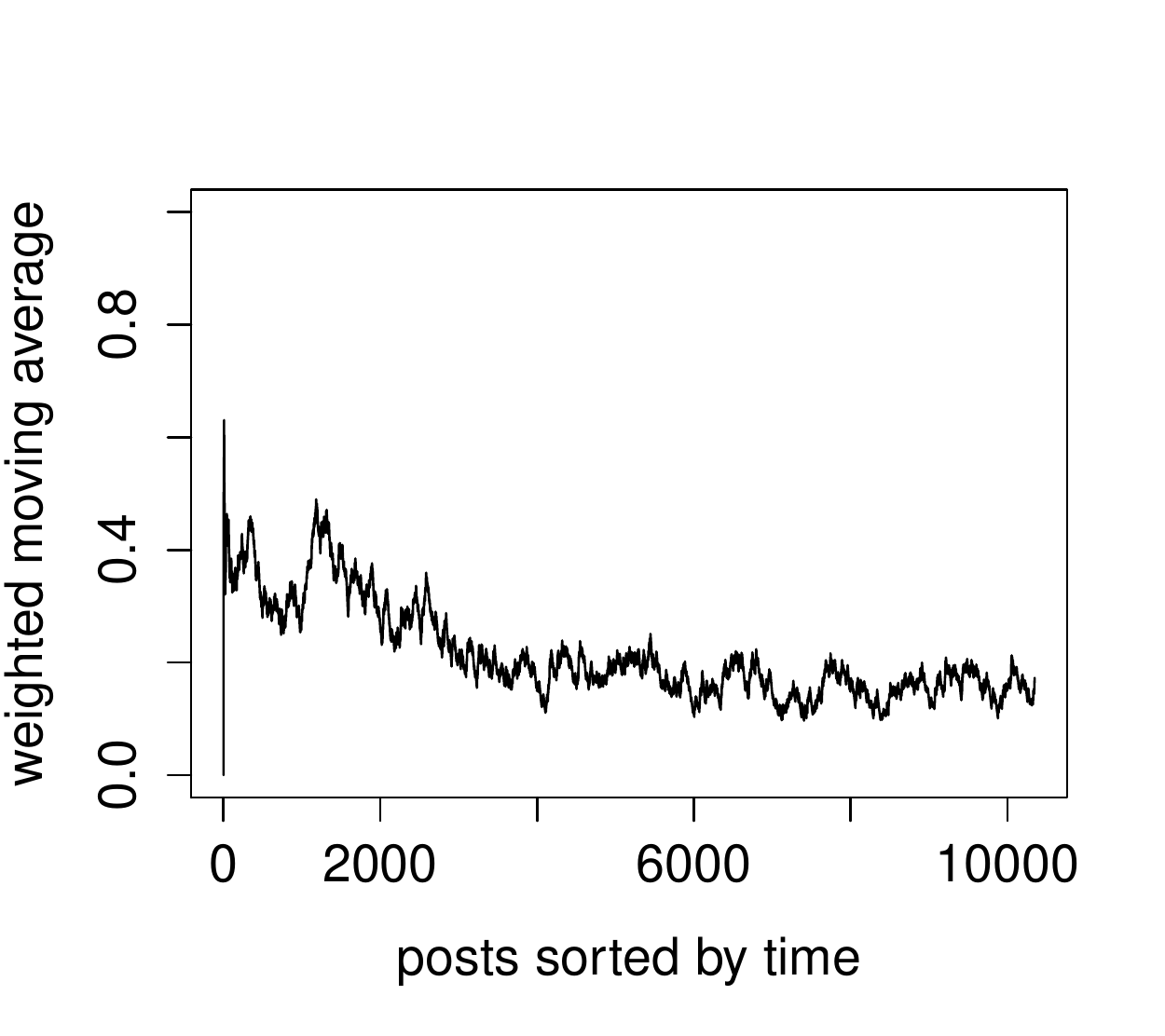}
\includegraphics[scale=0.34]{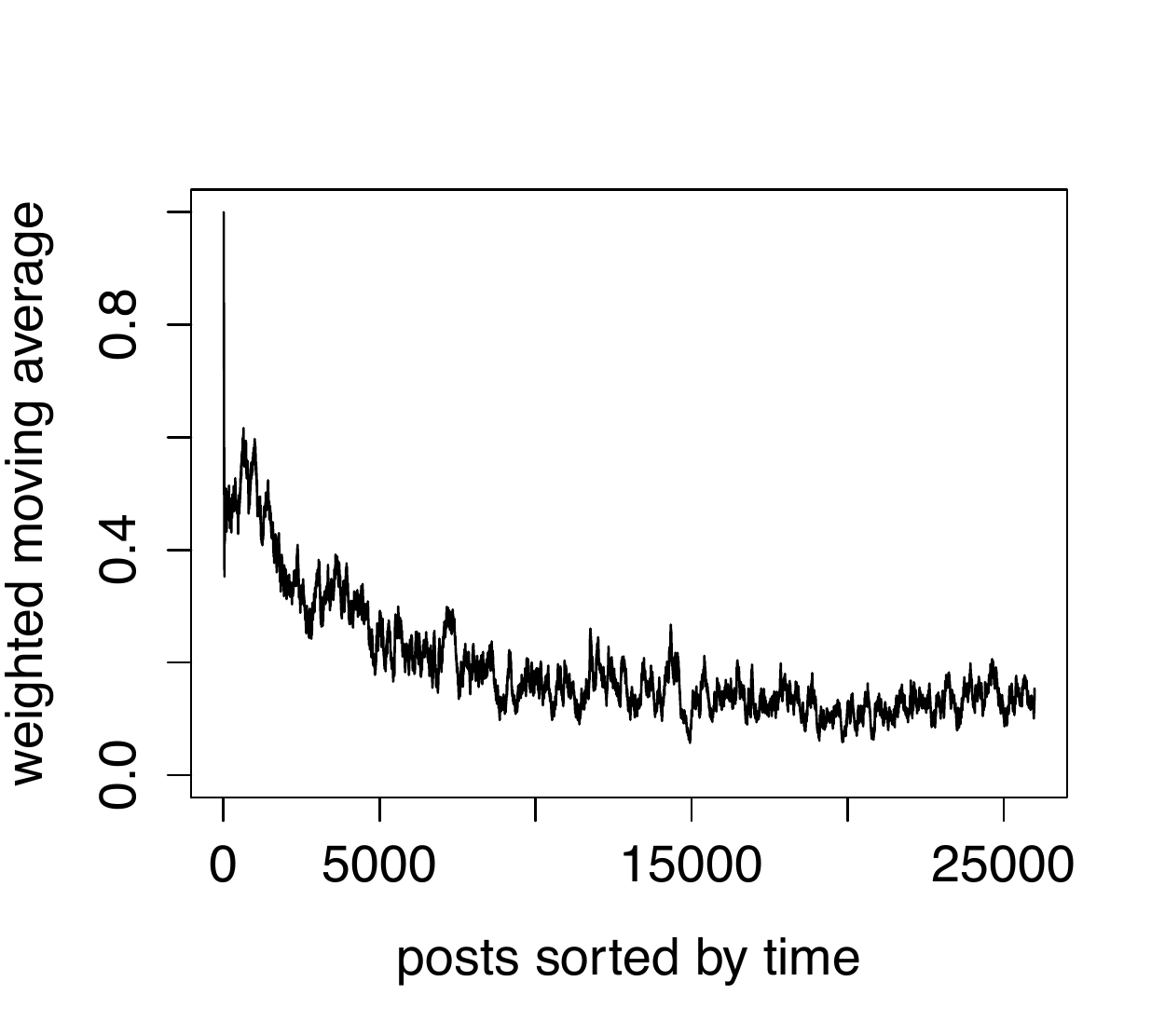}
\includegraphics[scale=0.34]{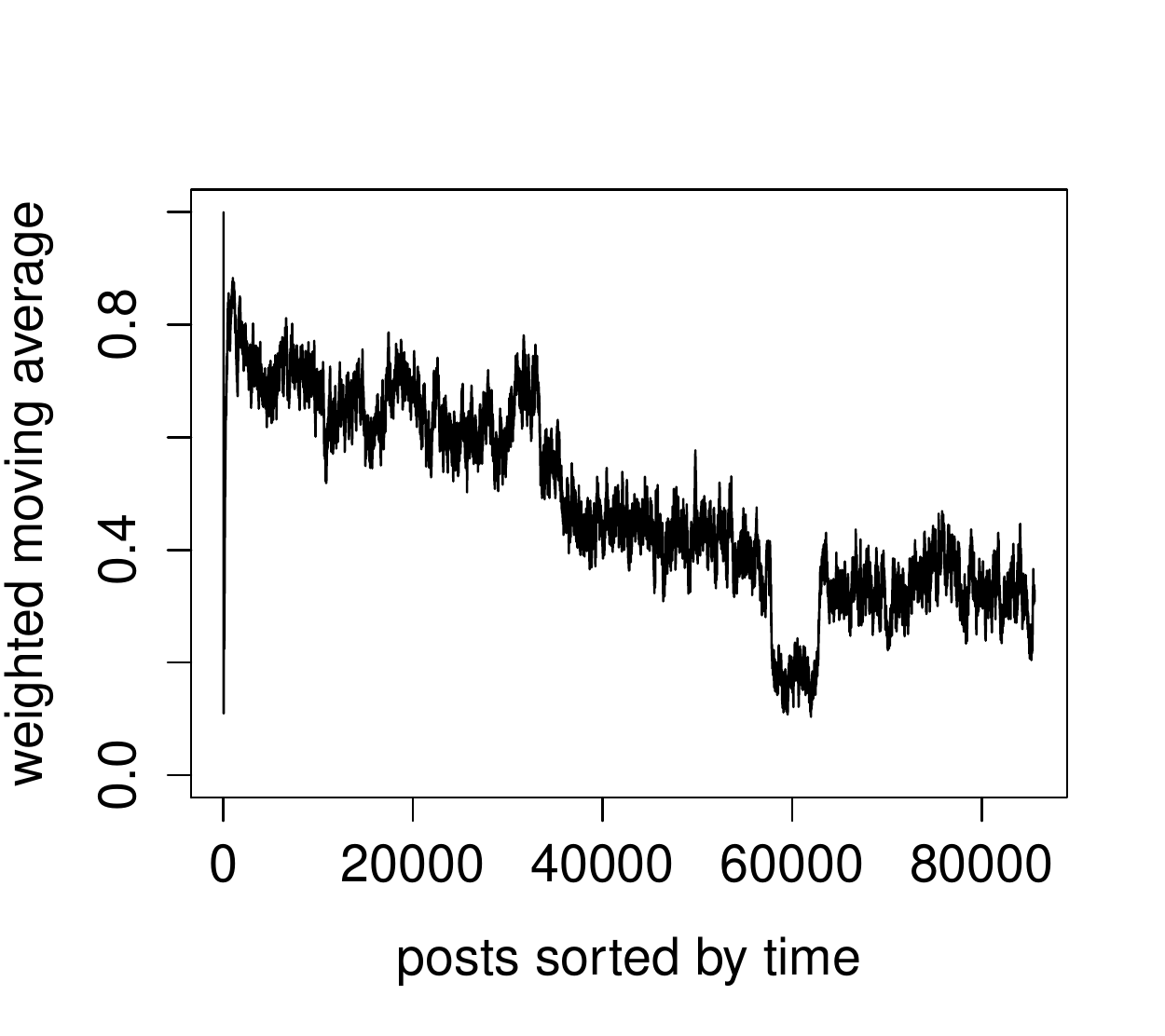}}
\end{center}
\vspace{-.5cm}
\caption{Statistics of small-talk by categories. Vocation: vocational courses. (Appl) Sci: Sciences and Applied Sciences. social-sci: Humanities and Social Sciences (Section~\ref{sec:prelim}).}
\label{fig:smalltalkclassify}
\end{figure*}

This paper studies both problems through a comprehensive dataset we crawled from the Coursera discussion forums. By examining the online discussion forums of all courses that were offered during the summer of 2013, we quickly discovered that both problems are ubiquitous in Coursera (see Fig.~\ref{fig:declines} and~\ref{fig:smalltalkclassify}). Thus, we believe it is natural to ask the following two questions:
\vspace{-.1cm}
{\quote{\emph{Question 1. How rapidly does the participation rate in the online discussion forums decline over time, and what behavioral factors maintain a robust participation rate?}}}
\vspace{-.1cm}
{\quote{\emph{Question 2. Is there a way to codify the generative process of  forum discussions into a simple model,
 and if so, can we also leverage our model to better facilitate users' navigation?}}}

\myparab{Motivation of these questions.}
The goal of studying Q1 is twofold. First, because online discussion forums are the main venue for teacher-student interaction and the only venue for student-student peer help, understanding the dynamics of the online discussion forums is crucial in understanding the quality of  students' online learning experience. Second, we believe  students' participation in  online forums is also linked to the fundamental problem of having high dropout rates in online education~\cite{WKO11}.  Exactly when (and why) the students
drop out of a course is not publicly accessible information. In fact, due to privacy concerns, this type of information may not be available in the near future. Furthermore, given the demographics of MOOC students and the nature of distance learning, it is unlikely for a student to pass a course without any participation  in the online forum discussion. The activeness of a course's online forum closely correlates with the volume of students that drop out of the course.

Crystalizing the formation of discussion threads into a simple model as asked in Q2 allows us to address the information overload problem and hence improve  students' online learning experience. Addressing information overload problems traditionally falls into the area of information retrieval (IR).  The primary goal here, however, is to highlight the unique characteristics of the dynamics of MOOCs' forums that are not seen in other online forums such as Yahoo! Q\&A and Stackoverflow (or other social media such as Twitter and Facebook). The generative model we develop in Section~\ref{sec:classify} guides us in choosing classifiers to filter out ``noise'' in discussions and in designing ranking algorithms to find the most course-relevant discussions.

Our goals in answering Q1 and Q2 are both to sustain forum activities and identify valuable discussions for individuals.

\myparab{Our methodology.} Our analysis consists of the following components.

\noindent{\emph{(1a) Statistical analysis.}} To address Q1, we carry out an in-depth analysis to understand the factors that are associated with students' participation in the online forums. Specifically, we first use regression models to understand what variables are significantly associated with the number of posts (or number of users) that appeared in the forum in each day for each course. One of the interesting discoveries, for example, is that the teaching staff's active participation in the discussion on average
increases the discussion volume \emph{but does not slow down the decline rate}. We also apply a standard $t$-test procedure to understand whether seeing too many new threads in a short time will reduce the depth of discussion in each thread. 
 Along the way we also present some basic statistics of Coursera,
such as the total number of students that participate in the forums, the distribution on the number of posts for each student, and the distribution on thread lengths.

\noindent{\emph{(1b) Identifying the information overflow problem.}} Based on the statistical analysis, one can see that  users have different needs in different stages of the course.
\begin{itemize}
\item In the first few days, the forum is often flooded with small-talk, such as self-introductions. The primary goal here is to \emph{classify} these small-talk threads and filter them out.
\item After that, the small-talk often starts to drop. At this point, most of the threads are valuable, so it is important to be able to give a relevance rank of new threads over time.
\end{itemize}
Thus, we need both a classifier for discussions and a relevance-ranking algorithm.
But we would like to understand a more fundamental question described in Q2: Is there a principled and unified way to consider the design of both the classifiers and the ranking algorithms. This question is addressed next.

\mypara{(2) Generative models.} We propose a \emph{unified generative model} for thread discussions
that simultaneously guides the (i)~choice of classifiers, (ii)~design of algorithms for extracting important topics in each forum, 
(iii)~design of a relevance ranking algorithm based on the topic extraction algorithm from (ii).  Our ranking algorithm is also compared with a number of baseline algorithms through \emph{human evaluation}.
Our simple model explains all the key experimental results we see.

\myparab{Related work.}
There has been a great deal of research in the areas of online social interaction and forum dynamics as well as a great deal of research devoted to online education. Here, we highlight a number of recent key works in these areas.

\mypara{MOOCs.}
Piech et al.~\cite{PiechHCDNK13} designed algorithms for a peer grading system to scale up evaluation of students. We are considering a different aspect of MOOC efficacy: attempting 
to foster engagement in discussion forums. Along these lines,  Ahn et al.~\cite{ahn2013} used a longitudinal fixed-effects model to identify the influencing factors of student participation in  peer-to-peer learning in a MOOC-type platform. Kizilcec et al.~\cite{Kizilcec2013} argued that student engagement in MOOCs should not be based solely on regular coursework completion. Cheng et al.~\cite{Cheng2013} designed tools to predict a student's risk of dropping out of the course.

Compared to the above works, our study is different in that: (1) It is based on a much more comprehensive dataset, 73 courses (almost all of the courses in Coursera during the summer 2013) versus at most 7 in previous work, (2) We identify new factors influencing engagement, and (3) it crystallizes the discussion dynamics via a generative model.

\mypara{Studies on forums.}
There are two main lines of research on the dynamics of online forums: (1) finding high-quality information or high-quality users in question-and-answer forums (see~\cite{Zhang2007ENO, Jurczyk2007,  Adamic2008,  Harper09, Anderson2012} and references therein), and (2) understanding social interactions in forums (\eg ~\cite{mendes2009socializing, Cha10measuringuser, SunCLWSZL11}). 
On the theoretical side, Jain et~al.~\cite{Jain2012} and Ghosh and Kleinberg~\cite{Ghosh2013} took a game-theoretic approach in understanding user behavior in online forums.  The work of Ghosh and Kleinberg is perhaps more relevant to our work, because they are interested in the optimal rate of instructor participation that encourages student participation.  

The online discussion forums in MOOCs differ from other social media studied in the following ways: 1. \emph{Both social and technical discussions are encouraged.} For example, in Stackoverflow~\cite{Anderson2012}, administrators aggressively remove low-quality threads, and in Twitter~\cite{Cha10measuringuser}, very few technical discussions occur. 2. \emph{Each forum focuses on one course.} Each course has one forum and only students in the course participate in the forum. A large portion of the students are also first-time users. While Yahoo!~Q\&A~\cite{Adamic2008,mendes2009socializing} has both social and technical discussions, MOOC forums have weaker social interaction and more focused technical discussions (and thus the techniques developed in~\cite{mendes2009socializing} are not directly applicable). 

\section{Preliminaries}\label{sec:prelim}

\myparab{Collecting the dataset. }
We focused on all 80 courses that were available in the middle of July and that ended before August 10, 2013.  Seven of these 80 courses became inaccessible while we were crawling or coding the data. Thus, we have complete records on 73 courses in total, which we used for our data analysis.  For developing the generative model, it is less important to have a complete forum data set, thus, we added four more courses that ended shortly after August 10, 2013. The dataset consists of approximately 830K posts (Section~\ref{sec:regression} presents more details). The full paper gives the entire list of courses. 

We first manually calculated  various course properties, such as the video length. Then, we  crawled the forum content, at a rate of 1 to 3 pages per second, from the Coursera server with Python and the Selenium library. Finally, we used Beautifulsoup to parse the data. 

\myparab{Categorizing the courses.} 
For the purpose of comparison across course types in this paper, we categorize a course as  \emph{quantitative} or non-quantitative, and \emph{vocational} or non-vocational. We adopt the following definitions: if a substantial portion of a course requires the students to carry out mathematical or statistical analysis, or to write computer programs, the course is a quantitative course. If a course's material is directly relevant to jobs that require high school or college degrees, or it is plausible to see the course offered in a typical university's continuing education division, then the course is considered a vocational course. Among these 73 courses, 37 of them are quantitative and 8 of them are vocational. There are 6 courses that are both quantitative and vocational. We partition the data in the following way for data summarization purposes: a course could be (1)~vocational, (2)~science or applied science (\ie~quantitative but not vocational) or (3)~humanities and social sciences (neither quantitative nor vocational).

\myparab{The structure of the forum.}  In a course forum, the students may create a new thread or add new content to an existing thread. Each thread consists of one or more ``posts,'' sorted in chronological order. The first post is always written by the creator of the thread. A user may respond to a thread (\ie adding a new \emph{post}) or respond to a post (\ie adding a new \emph{comment}).  In our analysis, we do not distinguish between posts and comments for the following two reasons: (1)~There is only a small portion of comments in the forum. (2)~The UI of the forums could be confusing like other forums, \ie the student could be unaware whether he/she is making a comment or adding a post. 

\myparab{Topics in a forum.}
The discussion threads can be roughly categorized into the following groups: (1) \emph{Small-talk} conversations that are not course specific, such as a self-introduction or requests to form study groups. (2) \emph{Course logistics} such as when to submit the homework, how to download the video lectures, \etc (3) \emph{Course-specific questions} that could be either very specific or open-ended. 

Among these three categories, the last one is the most valuable to the students' learning process. The second category could be quite relevant, especially if a substantial portion of the threads are about homework and exams. The first one, however, can be quite disruptive to many students when the number of new threads is already excessive.
Thus, we want to understand how many small-talk posts there are for each course and whether the portion of small-talk we see changes over time.  

We answer the first question with the help of Amazon Mechanical Turk (MTurk). Specifically, we randomly choose 30 threads from each course and hire workers from MTurk to label the threads. Each thread is labeled by 3 people and we take a majority vote to determine the labels. Figure~\ref{fig:smalltalkclassify}a shows the distributions of the small-talk by category. We can see that a substantial portion of the courses have more than 10\% of small-talk in the online discussion forums. Furthermore, the humanities and social sciences courses have a higher portion of small-talk.

\myparab{Temporal dynamics of small-talk.} Now we study how the portion of small-talk changes over time. Since it is infeasible (in time and cost) to label a significant portion of the threads, we use a machine learning approach (specifically, a support vector machine, see Section~\ref{sec:classify}) to classify the threads by using the training data labeled through MTurk. We put all threads under the same category together. Then we sort the threads by the time elapsed between the beginning of the course and the creation of the thread. Here we focus on only the threads created within 35 days after the class started. Then we compute the ``moving average'' as follows: let $h_1, h_2, ..., h_m$ be the sorted threads within the same course and $\eta_i$ be an indicator variable that sets to 1 if and only if $h_i$ is classified as small-talk. Then we define our moving average at time $t$ as:
$s_t = \frac{\sum_{1 \leq i \leq t}\eta_i \alpha^{t - i}}{\sum_{1 \leq i \leq t}\alpha^i}.$
Figure~\ref{fig:smalltalkclassify} ($\alpha=.99$) shows the results. We can see that at the beginning, the percentage of small-talk is high across different categories, and then it drops over time. However, 
for humanities and social sciences courses, on average more than 30\% of the threads are classified as small-talk even long after a course is launched. 

We remark that these plots  give just estimates on the volume of small-talk. There are two types of noise presented in these plots: (1)~We are aggregating all the threads in the same category, so course-level information could be lost. (2)~The support vector machine could have classification errors. 
 
Nevertheless, we may conclude that small-talk is a major source of information overload in the forums. 

\myparab{Why the existing infrastructure is insufficient.} We notice that the Coursera platform allows the instructors to categorize the threads. However these categories are customized by the teaching staff. Some categorizations are more effective than others. And there is no effective mechanism to force the students to respect the categorization consistently. It would be infeasible for the staff to manually correct the labels when the new threads flood into the forum.

\section{Statistical analysis}\label{sec:regression}
This section examines the following 
two subquestions of Q1: (1a) What factors are associated with the decline rate of the online forum participation?
(1b) Will having more discussion threads in a short period of time dilute students' attention?

We use linear regression to answer 1a and student's $t$-test to answer 1b.

\subsection{Analysis of forum activity decline}
This section studies what factors are associated with the decline rate of  forum activities. Here, our dependent variables are $y_{i,t}$ and $z_{i,t}$, where $y_{i,t}$ refers to the \emph{number of posts} at the $t$-th day in the $i$-th course, and $z_{i,t}$ refers to the number of \emph{distinct users} that participate in the discussion in the $t$-th day in the $i$-th course. Both variables are important to the students' online, social learning experience.

The following variables could be relevant to $y_{i,t}$ and $z_{i,t}$:

\myparab{Quantitative ($Q_i$)} is an indicator (boolean) variable that sets to $1$ if and only if the $i$-th course is a quantitative course.

\myparab{Vocational ($V_i$)} is an indicator variable that sets to $1$ if and only if the $i$-th course is a vocational one.

\myparab{Total length of the videos ($L_i$)} is the sum of the length of all lecture videos in the $i$-th course (in hours).

\myparab{Total duration ($D_i$)} is the total length (in days) of the $i$-th course.

\myparab{Peer-graded ($P_i$)} is an indicator variable that sets to $1$ if and only if one or more assignments are reviewed/graded by  peer students.

\myparab{Teaching staff's activeness ($S_i$)} is the number of posts the teaching staff make throughout the $i$-th course.

\myparab{Graded homework ($H_i$)} is the total number of homework assignments that are graded by the teaching staff.

 \myparab{Intrinsic popularity ($M_i$, or $M'_i$)} is the volume of  forum discussion in the beginning of the course, defined as $M_i$, the median number of posts (if our dependent variables are $y_{i,t}$), or $M'_i$, the number of distinct users (if our dependent variables are $z_{i,t}$) in the first three days. Roughly speaking, this variable captures the ``intrinsic popularity'' of each course, \eg it is likely that a course on public speaking  is more popular than a typical course in electrical engineering.

We  start by walking through the behavior of the variables above before we present and analyze our model.

\vspace{-.2cm}
\subsubsection{Statistics of Coursera}
We now examine the key statistics for the 73 courses  collected in Coursera, starting with our ``dependent variable,'' the behavior of the students.

\begin{figure}
\begin{center}
\subfigure[\# of posts per student.]{
\includegraphics[scale=0.3]{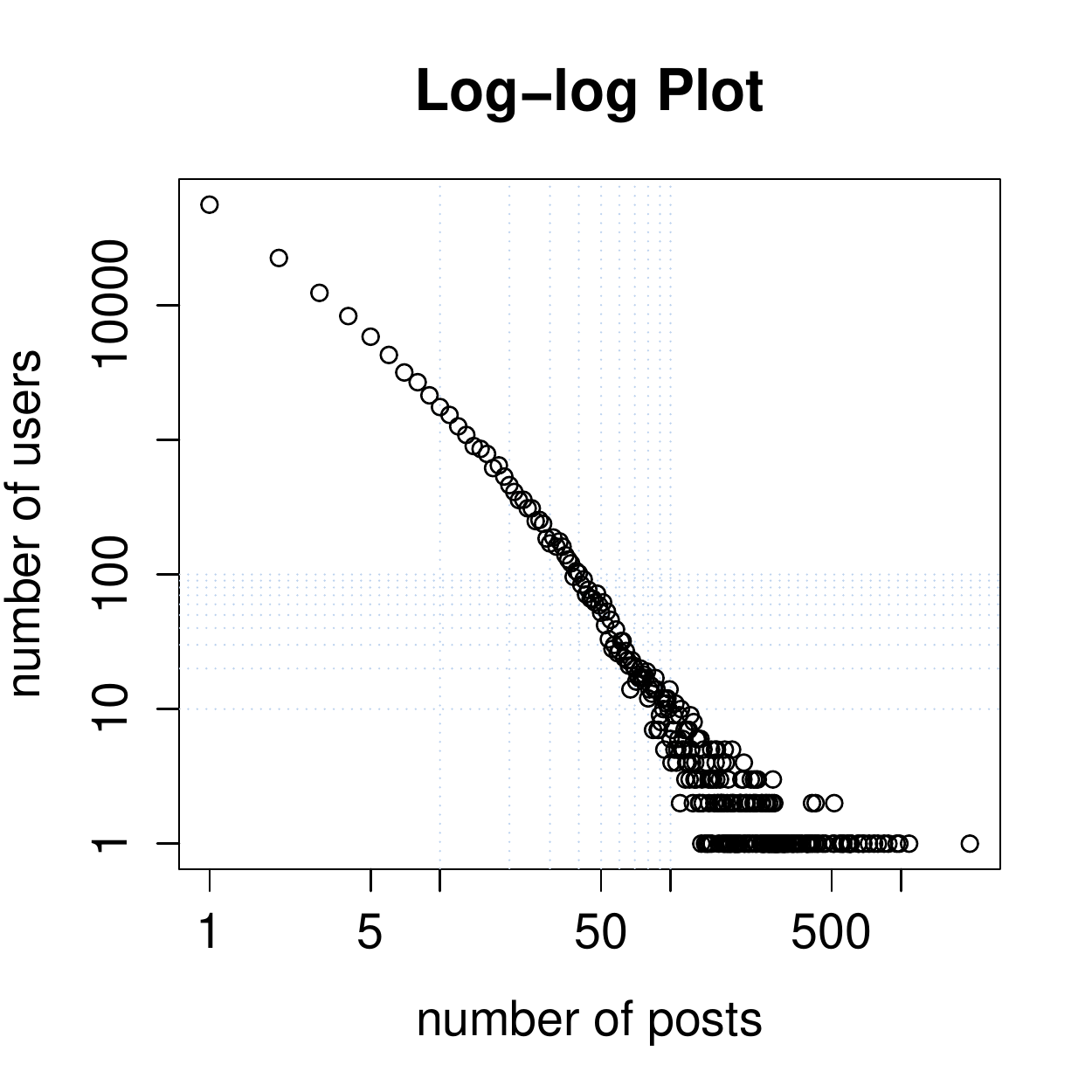}}
\subfigure[CDF of decline rate.]{
\includegraphics[scale=0.3]{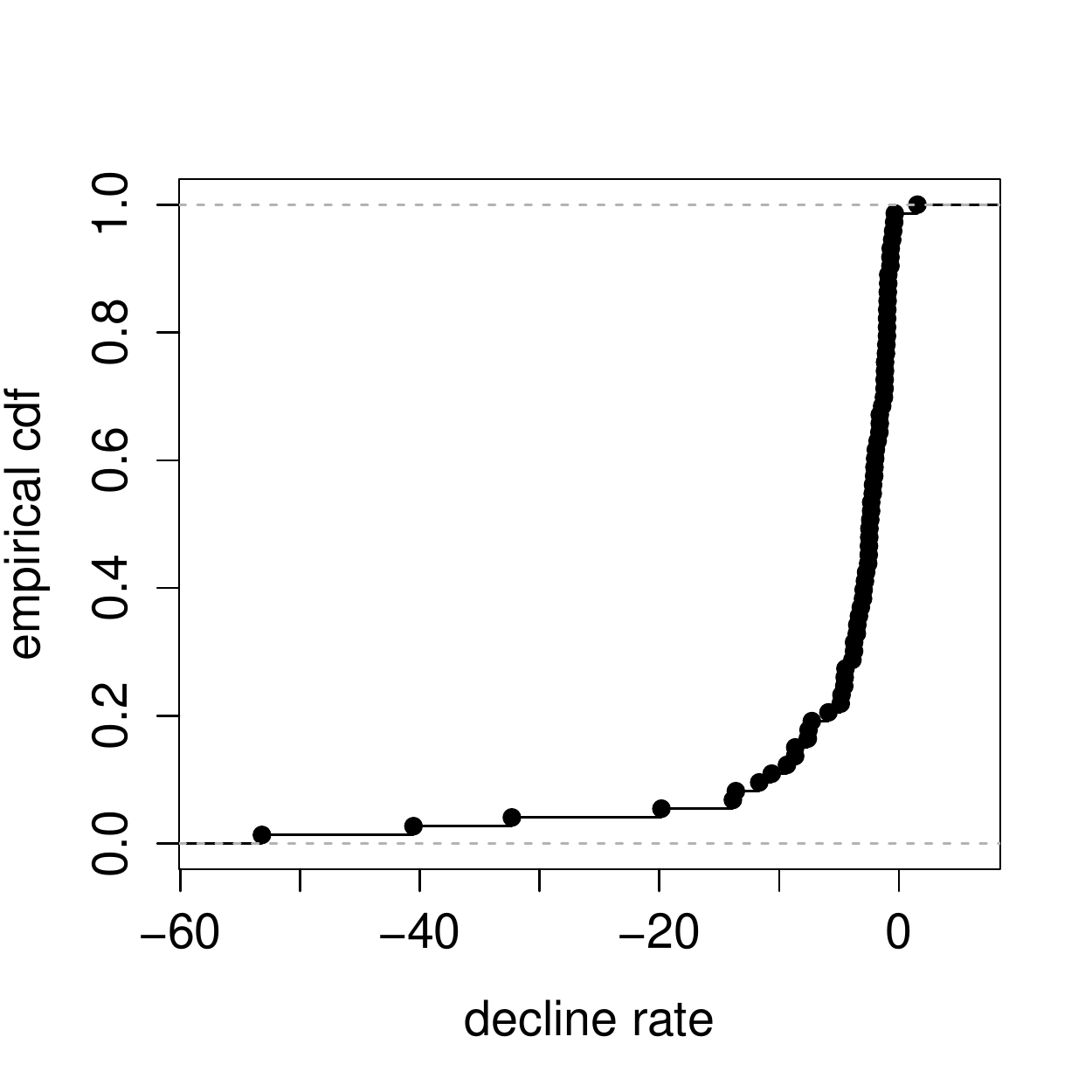}}
\end{center}
\vspace{-.5cm}
\caption{Distribution of student posts and decline rates.}
\label{fig:postdist}
\end{figure}

\myparab{Students in the forums.}
In these 73 courses, there are 171,197 threads,  831,576  posts, and 115,922 distinct students. Figure~\ref{fig:postdist}a shows the distribution of the number of posts each student made (in  log-log scale).

\begin{figure}
\begin{center}
\vspace{-.5cm}
\includegraphics[scale=0.3]{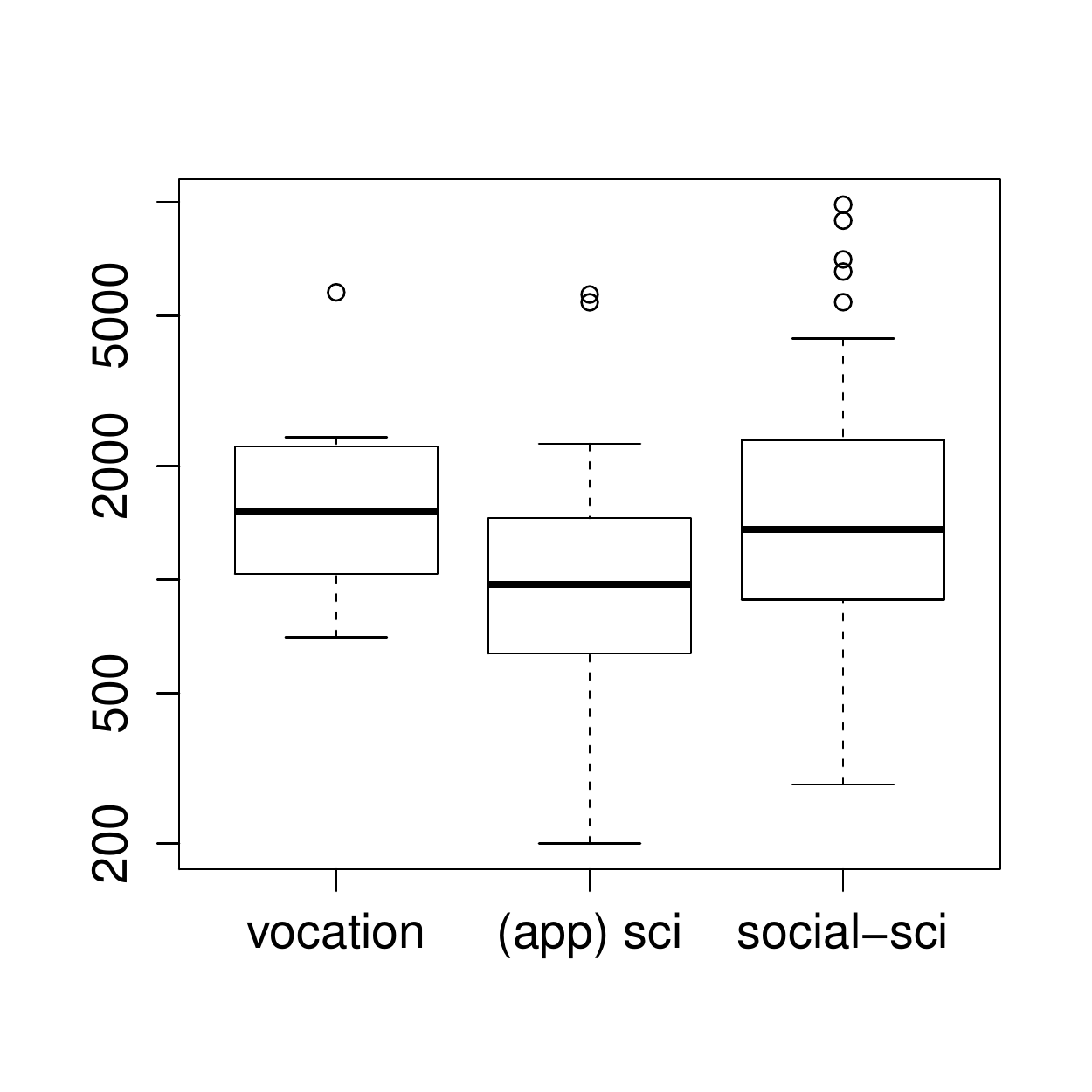}
\includegraphics[scale=0.3]{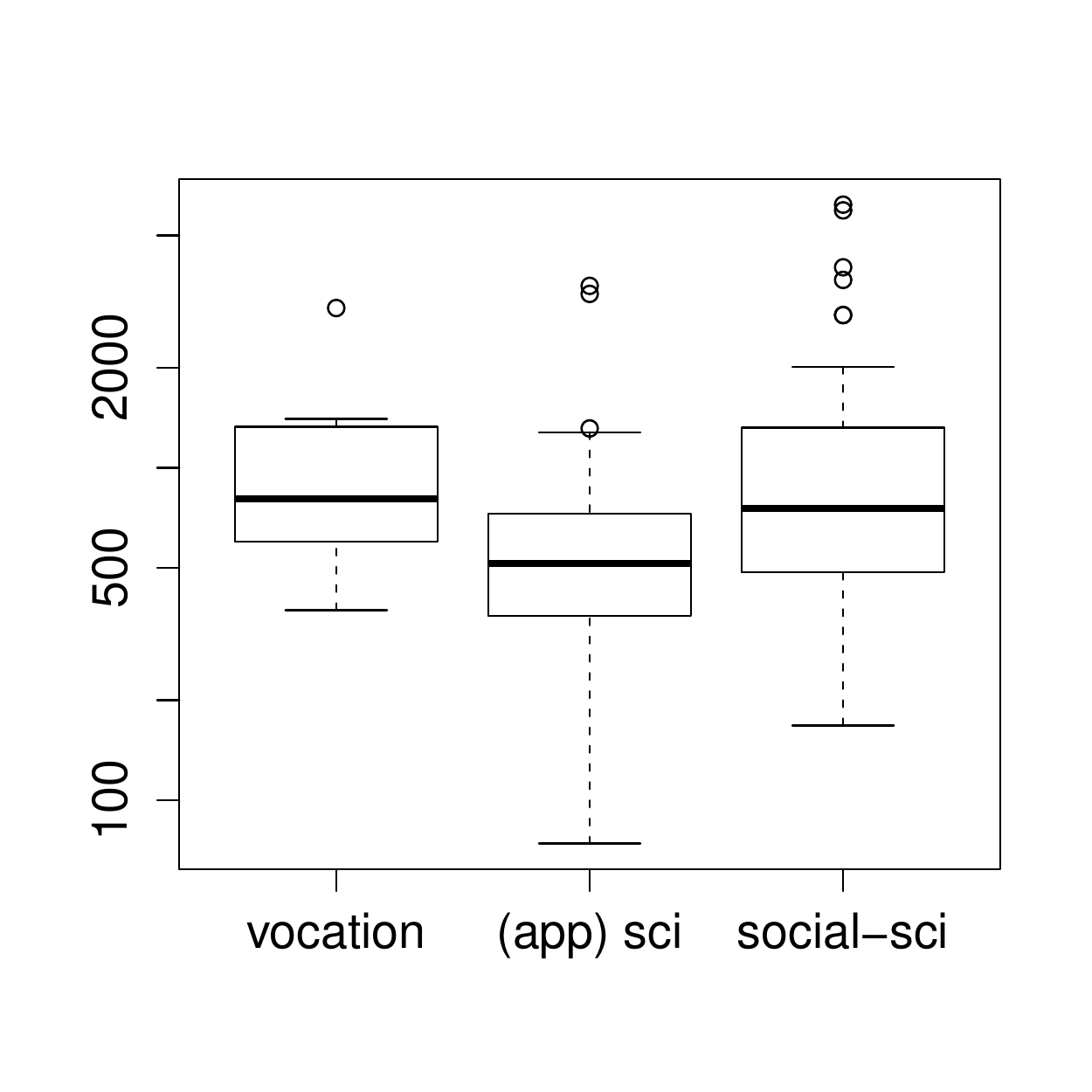}
\end{center}
\vspace{-.5cm}
\caption{Distribution of the number of students (left) and active students (right) per course.}
\label{fig:cdfstudent}
\end{figure}

We may roughly classify the students in the forum as ``active'' and ``inactive.''  Active students refer to those who make at least 2 posts in a course.\footnote{Choosing $2$ as the threshold is quite arbitrary. The goal here is to show that many students make a small number of posts.} Inactive students are everyone else. Figure~\ref{fig:cdfstudent} shows the distribution of the number of students/active students across different courses by category. 
Overall, the average number of students per course is 1835.0 (sd $= 1975.4$) but this number reduces to 
significantly 1069.7 (sd = $1217.7$)
for active students.

\myparab{Distribution of decline rate.} 
We next present how $y_{i,t}$ and $z_{i,t}$ change over time for different courses. Figure~\ref{fig:declines}a shows the variables $\{y_{i,t}\}_{t \geq 1}$ for five randomly selected courses. We also fit linear models on each of these datasets, where the regressors are the index of the day. All the courses presented here experienced decline in participation. The rest of the courses also behaved similarly.  Figure~\ref{fig:postdist}b shows the distribution of the decline rate for all the courses (mean $= -5.0$ and sd = $8.7$; 72 of 73 courses are negative\footnote{The only positive one also has a negative decline rate when we count the \# of distinct users.}). 
The variables $\{z_{i,t}\}_{t\geq 1}$ are qualitatively similar. We next study the distribution on the count differences between two consecutive days in the same course. It is not uncommon to see outliers due to large fluctuations in discussions, especially in the beginning of the course or when homework/exams are due. But after we remove the top and bottom 3\% outliers in each course, the count differences follow a Gaussian distribution for most of the courses (see Figure~\ref{fig:declines}b for the Q-Q plots). These five courses all pass Shapiro's normality test ($p$-values $\geq 0.01$). Overall, 51  of 73 courses have $\geq 0.01$ $p$-value in Shapiro's test.

\myparab{Video length.} Among these 73 courses, the mean video length is 12.71 hours  (sd $=7.85$). The distribution fits quite well with Gaussian. We do not see discrepancies in distributions across categories.

\myparab{Length of the courses.} All the courses are between 4  and 14 weeks long. The mean is 58.8 days (sd $=15.3$).

\myparab{Homework assignments.}
The mean number of staff-graded assignments per course is 10.13 (sd $= 10.88$).Out of the 73 courses, 6 of them do not have any staff-graded assignments. As for peer-graded assignments, there are a total number of 39 courses that
have peer-graded homework. Among them 5 courses are vocational, 11 courses are quantitative but not vocational, 23 of them are from the rest of the courses.

\myparab{Teaching staff activity.} On average, there are 366.9 posts in each course made by the teaching staff (sd = $446.1$). But there are two courses that have no posts from the teaching staff.

\myparab{Postulation of a model.} Based on the evidence presented above, we postulate the following linear model on the post counts: let $y_{i,t}$ be the number of posts on the $t$-th day in the $i$-th course. We assume $y_{i,t+1} - y_{i,t} \sim N(\mu_i, \sigma_i)$, \ie $y_{i, t} = \sum_{j \leq t}N(\mu_i, \sigma_i) = N(t \mu_i, \sqrt t \sigma_i)$. The mean term grows linear in $t$ while the ``noise term'' grows linear in $\sqrt{t}$. When $t$ is sufficiently large, the mean term dominates $y_{i,t}$. In other words, we may model $y_{i,t} = A_it + B_i + \epsilon_{i,t}$, where $A_i$ and $B_i$ only depend on the factors of the $i$-th course. Note that while serial dependency can be present, we believe this factor-adjusted, deterministic, linear trend is sufficient to explain the pattern; this is confirmed by our subsequent empirical results.

\subsubsection{Regression model}
We now present our linear model. More concretely, $y_{i, t}$ is linearly related to the course factors, the variable $t$, and the interacting terms between $t$ and the factors, as
\begin{equation}\label{eqn:simplify}
y_{i,t} =   At + B,
\end{equation}
where $A = \beta_1 Q_i + \beta_2 V_i + \beta_3 L_i  + \beta_4 D_i + \beta_5 P_i + \beta_6 S_i + \beta_7 H_i + \beta_8 M_i + \beta_{16}$ and $B = \beta_0 + \beta_9 Q_i + \beta_{10}V_i + \beta_{11}L_i  + \beta_{12}D_i + \beta_{13}P_i + \beta_{14} S_i + \beta_{15}M_i$.
Thus, we can view $\beta_1, ..., \beta_8$, and $\beta_{16}$ as influencing the \emph{decline rate} of forum participation whereas $\beta_0$, $\beta_9$, ..., $\beta_{15}$ are influencing the initial participation volume.

\begin{table}
\begin{center}
\small
\begin{tabular}{l D{.}{.}{5.5} @{}D{.}{.}{5.5} @{}D{.}{.}{5.5} @{}}
\toprule
            & \multicolumn{1}{c}{On $y_{i,t}$} & \multicolumn{1}{c}{On $z_{i,t}$} & \multicolumn{1}{c}{On $\log(z_{i,t})$} \\
\midrule
(Intercept) & 18.276         & 70.252^{***}  & 4.268^{***}  \\
$Q_it$         & 1.511^{***}    & 0.847^{***}   & 0.014^{***}  \\
$V_it$         & 3.328^{***}    & 1.463^{***}   & 0.011^{**}   \\
$L_it$         & -0.071^{***}   & -0.024^{***}  &              \\
$D_it$         & 0.034^{***}    & 0.025^{***}   & 0.001^{***}  \\
$P_it$         & -0.631^{**}    & -0.375^{***}  & 0.003        \\
$S_it$        & -0.168^{**}    & -0.067^{**}   & -0.001^{**}  \\
$H_it$        & 0.000          & -0.001        &              \\
$M_it$ (or $M'_it$)        & -0.007^{***}   & -0.005^{***}  & 0.000^{**}   \\
$Q_i$          & -13.975        & -23.737^{***} & -0.185^{**}  \\
$V_i$          & -135.567^{***} & -61.404^{***} & -0.153       \\
$L_i$         & 1.960^{**}     & -0.049        &              \\
$D_i$          & -0.561         & -0.624^{***}  & -0.010^{***} \\
$P_i$          & 88.289^{***}   & 32.005^{***}  & 0.247^{***}  \\
$S_i$          & 6.050^{**}     & 3.249^{***}   & 0.074^{***}  \\
$H_i$          & 1.398^{**}     & 0.973^{***}   &              \\
$M_i$  (or $M'_i$)       & 0.481^{***}    & 0.360^{***}   & 0.003^{***}  \\
$t$           & -1.864^{***}   & -1.980^{***}  & -0.071^{***} \\
\midrule
R$^2$       & 0.555          & 0.467         & 0.530        \\
Adj. R$^2$  & 0.554          & 0.465         & 0.526        \\
Num. obs.   & 5074           & 5074          & 1711         \\
\bottomrule
\vspace{-2mm}\\
\multicolumn{4}{l}{\textsuperscript{***}$p<0.01$, 
  \textsuperscript{**}$p<0.05$, 
  \textsuperscript{*}$p<0.1$}
\end{tabular}
\normalsize
\end{center}
\vspace{-.2cm}
\caption{Models of forum activities}
\vspace{-.3cm}
\label{table:forum}
\end{table}

The first column in Table~\ref{table:forum} shows the results based on ordinary least-squares regression.
We make a number of observations. First, it is evident that $y_{i,t}$ correlates well with the intrinsic popularity $M_i$, but the magnitude of impact of $M_i$ on the decline rate in the long run appears very light. The coefficients
of $Q_i, V_i, Q_it$, and $V_it$ suggest that while being a quantitative/vocational course will initially reduce the volume of forum participation, in the long run the course also experiences a lower decline rate (all the $p$-values $\leq 10^{-6}$).

The coefficients associated with $S_i$ also look quite surprising: while the teaching staff's active participation in the forum correlates with the increased volume of discussion (the addition of one post made by the teaching staff corresponds to an increase of 6.05 posts \emph{per day}), in the long run  their participation does not seem to reduce the decline rate. In fact, there is evidence that an increase in staff  participation correlates with a \emph{higher} decline rate ($p$-value $= 0.021$).

The variable $P_i$ (whether the course has peer-reviewed homework) behaves in a similar manner: while having peer-reviewed homework adds 88.29 posts per day on average, it also moderately increases the decline rate ($p$-value $=0.018$).

Finally, the $p$-value of the model is $2.2\times 10^{-16}$ , suggesting overall significance. We also diagnose the residuals, which do not seem to elicit any heteroscedastic pattern. We further check the differences between the slope of $t$, \ie~the quantity $A_i$, under our proposed regression model, and the counterpart for regression only on $t$ for each course; we observe that these differences are also reasonably small in magnitude.

We next move to the model for the variables $z_{i,t}$. We use the same set of regressors except that we substitute $M_i$ with $M'_i$. Specifically, we use the following model:
{\small
\begin{eqnarray*}
z_{i, t} & = & \beta_0 + \beta_1 Q_it + \beta_2 V_it + \beta_3 L_i t + \beta_4 D_i t + \beta_5 P_it \\
 & & \quad + \beta_6 S_it + \beta_7 H_it + \beta_8 M'_it +  \beta_9 Q_i + \beta_{10}V_i + \beta_{11}L_i \\
& & \quad + \beta_{12}D_i + \beta_{13}P_i + \beta_{14} S_i + \beta_{15}M'_i + \beta_{16}t.
\end{eqnarray*}}
\vspace{-.2cm}

Our results are shown in the second column of Table~\ref{table:forum}. We can see that qualitatively, the variables correlate with
$z_{i,t}$'s in a similar manner: (1) Quantitative and vocational courses attract a smaller volume of discussion in the beginning but they also correlate with smaller decline rates (all with $p$-value $\leq 10^{-6}$). (2) $S_i$ correlates with an increased number of distinct users ($p$-value $= 0.00994$) but there is evidence that it correlates with  higher decline rates ($p$-value $= 0.038$). (3) $P_i = 1$ increases the total number of distinct users ($p$-value $=  5.94 \times 10^{-10}$) and it also correlates with  higher dropout rates ($p$-value $= 0.0016$).  Finally, the $p$-value of the model is $2.2\times 10^{-16}$, suggesting again an overall significance of the model, and the residuals do not show any obvious pattern.


\myparab{More robust linear model on a subset of courses.}
While the linear regressions are in general quite robust against noises, we would also like to restrict ourselves to those courses that have ``nice'' behavior so that the residuals exhibit normality, and determine whether the conclusions are consistent. Specifically, we choose
those 51 courses whose count differences in posts (after removing the top and bottom 3\% outliers) pass the Shapiro test (with $p$-value $\geq 0.01$; see discussion in Section~\ref{sec:prelim}).

We fit the data from these 51 courses with the following model:
\begin{eqnarray*}
\log (z_{i,t}) & = & \beta_0 + \beta_1 Q_it + \beta_2 V_it + \beta_3 D_it + \beta_4 P_it + \beta_5 S_it \\
& &   + \beta_6 M'_it  + \beta_7 Q_i + \beta_8 V_i + \beta_9 D_i + \beta_{10}P_i  \\
& &  + \beta_{11}S_i + \beta_{12}M'_i + \beta_{13}t.
\end{eqnarray*}
The logarithmic transformation is performed because then we observe a higher significance of the model (but this is not strictly necessary and the result otherwise is similar). The variables $L_i$, $L_it$, $H_i$, and $H_it$ are not statistically significant so they are removed from the model. The third column of Table~\ref{table:forum} presents our results.

We also tested the normality of the residuals.
The $p$-value of the Shapiro test is $0.148$, which indicates that our model fits quite well for these 51 courses. 
The conclusions we see here are mostly consistent with the conclusions made from the above linear models.

\subsection{Attention to each thread}
Next, we investigate the following question: ``Will having more threads created at the same time reduce
the average ``attention'' each thread receives on average?''

\begin{figure}
\begin{center}
\subfigure[Log-log plot of  thread length distribution]{
\includegraphics[scale=0.3]{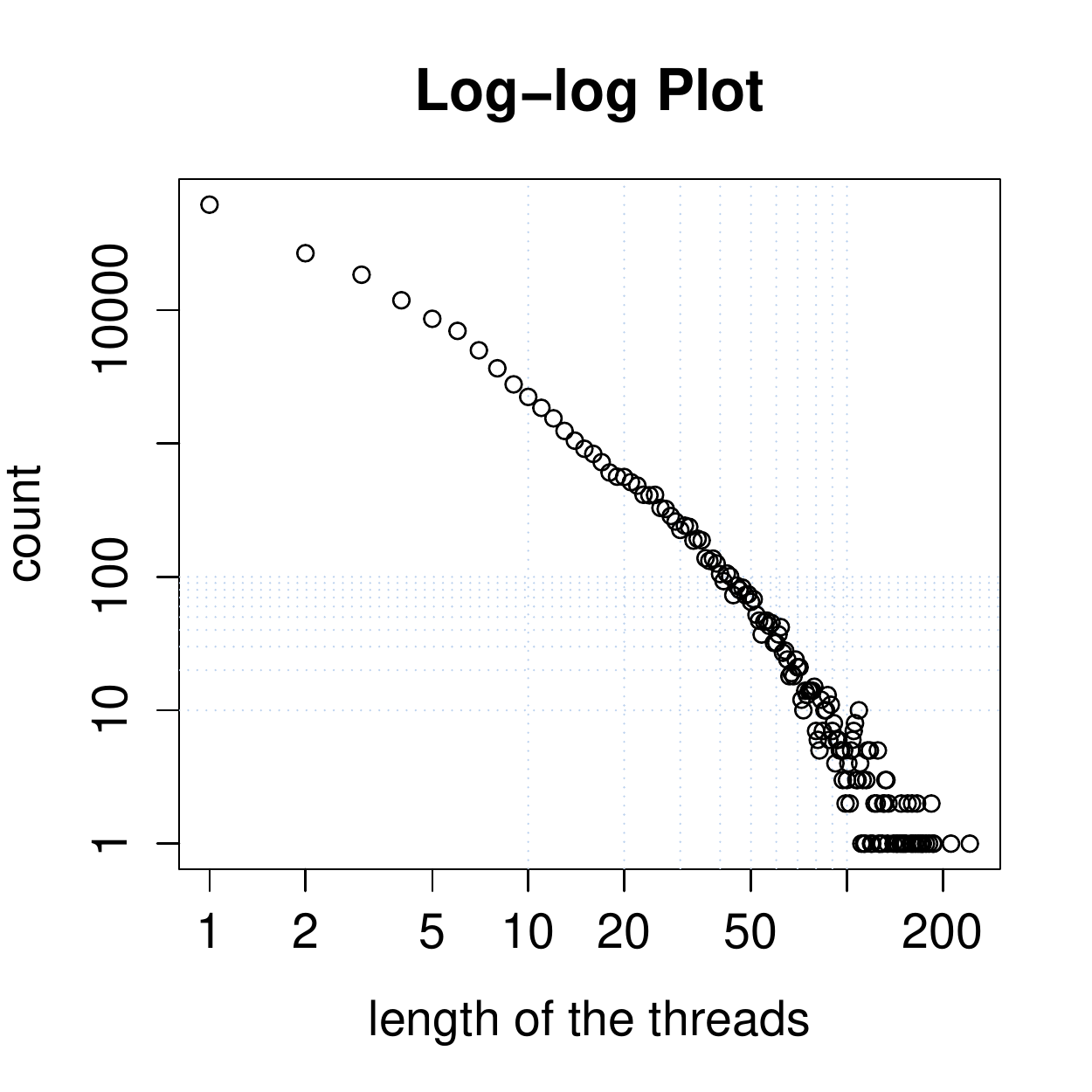}}
\subfigure[Threads' length by category]{
\includegraphics[scale=0.3]{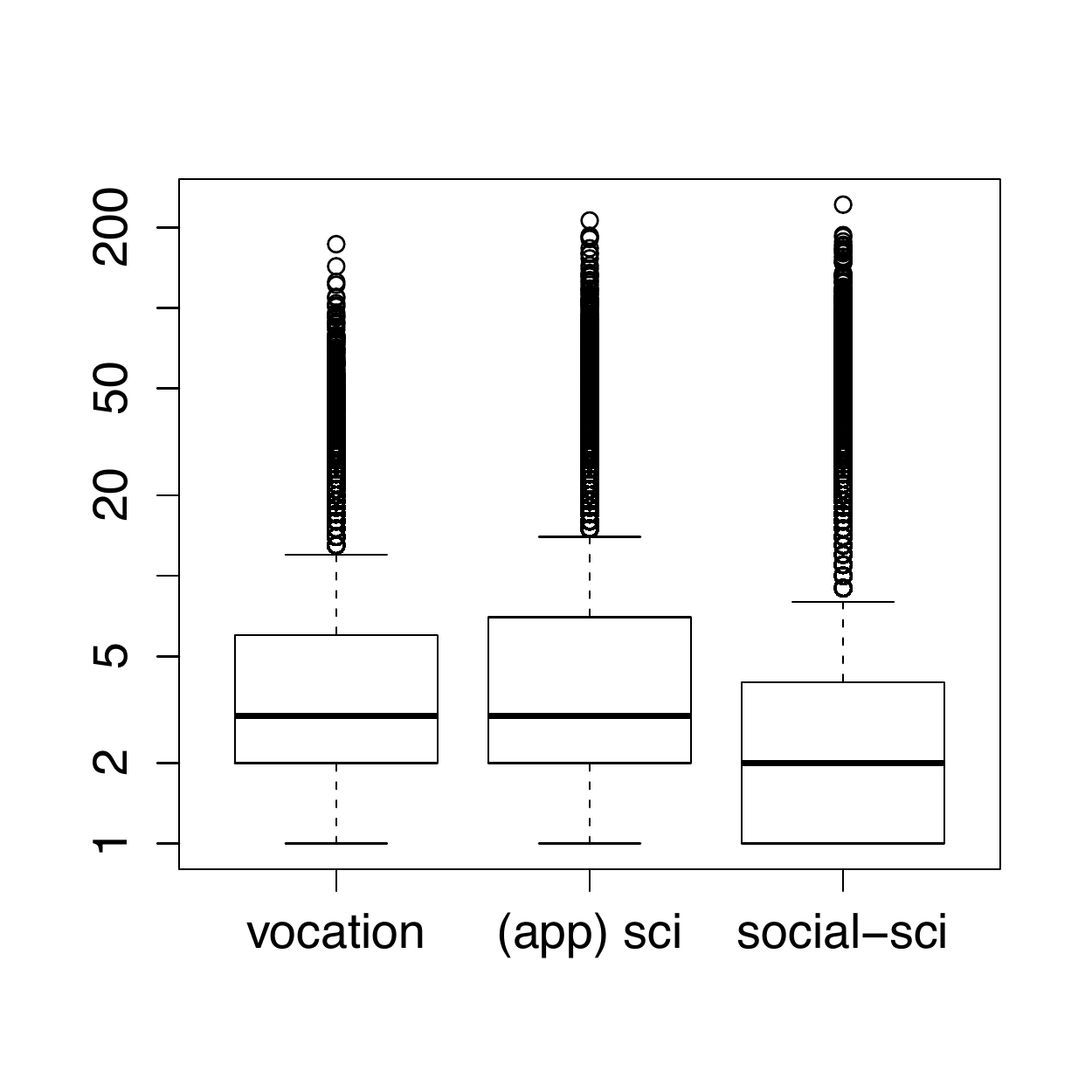}}
\end{center}
\vspace{-.4cm}
\caption{Distribution of thread lengths.}
\label{fig:threaddistr}
\end{figure}

\myparab{Distribution of the thread length.}
Let us start with understanding the distribution of the length of the threads in the discussion forum.
The mean of the length of \emph{all  threads} is 4.98 (median $=2$ and sd $= 8.65$). Figure~\ref{fig:threaddistr}a gives the distribution of thread length in log-log scale. Figure~\ref{fig:threaddistr}b also gives the boxplots of the distribution of the threads' length by categories.

\myparab{Independent two-sample $t$-tests.} We use a $t$-test to understand whether having more newly created threads will dilute the attention
each thread receives.
Let $h_{i}$ be a discussion thread in the forum and let $\ell_i$ be the length of the thread (the total number of posts and comments in the thread). Also, let $f(h_i, t)$ be the total number of other threads that were created within $t$ days before and after $h_i$ is created in the same course. For example, if $h_i$ was created on July 2nd 3pm, then $f(h_i, 1)$ is the number of threads other than $h_i$ that were created between July 1st 3pm and July 3rd 3pm.

\begin{figure}
\begin{center}
\subfigure[Length of threads vs. \# of new threads]
{

\includegraphics[scale=0.24]{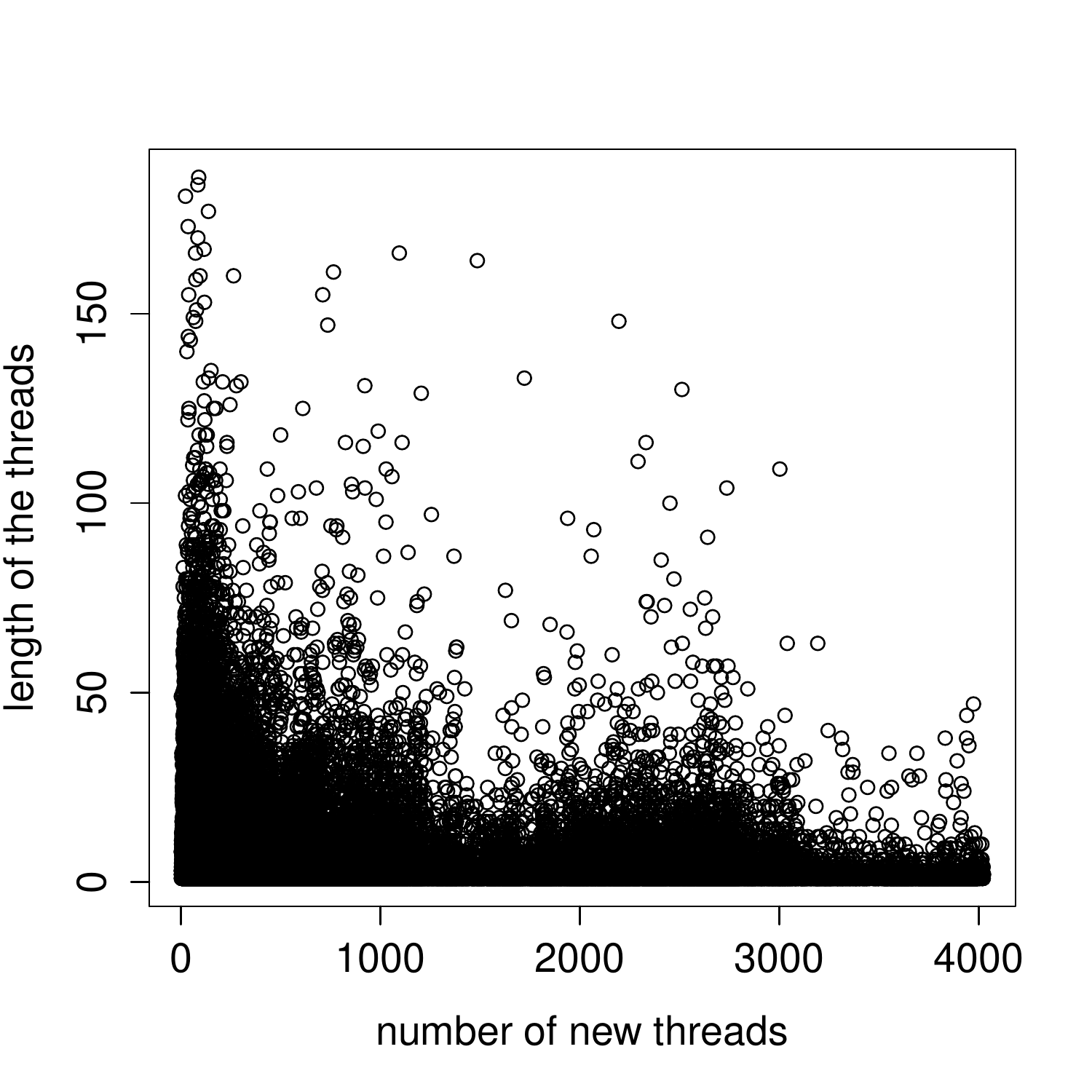}

}
\quad
\subfigure[Boxplots of two groups ($\leq 140$ new threads and otherwise)]{
\includegraphics[scale=0.3]{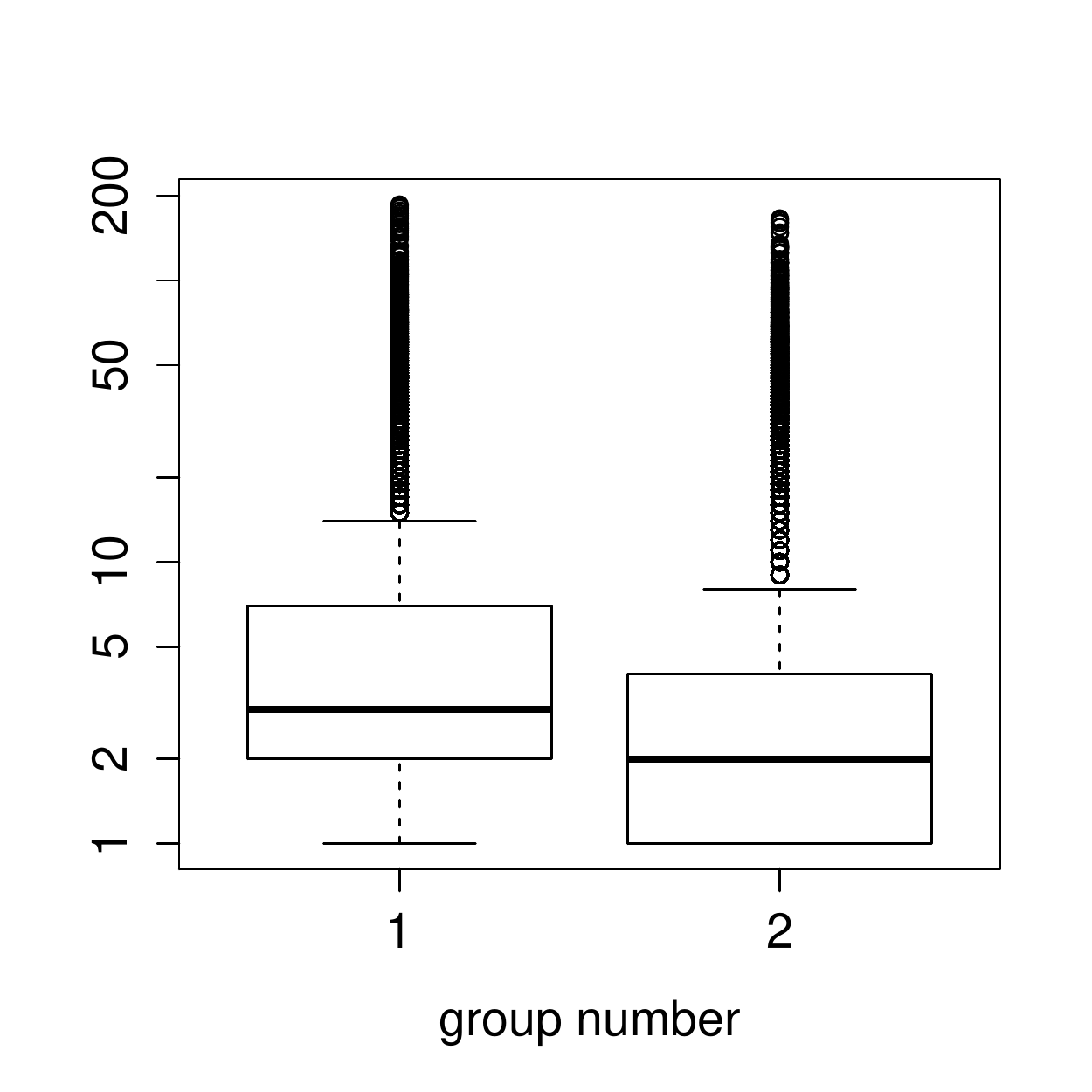}}
\end{center}
\vspace{-.3cm}
\caption{The length of discussion threads vs. the number of threads created around the same time.}
\label{fig:lenthreads}
\end{figure}

Figure~\ref{fig:lenthreads}a shows the plot of $\ell_i$ against $f(h_i, 1)$. We could fit this data set with a linear model. But we find the linear model's explanatory power is quite low (\ie $R^2$ is below 0.02). Thus, we resort to two-sample procedures to analyze this group of data set. Specifically, we partition the threads into two groups. The first group $G_1$ contains all the threads $h_i$ such that $f(h_i, 1) \leq 140$. The second group $G_2$ contains the rest of them. The threshold is chosen in a way that both the size and $\ell_i$'s variances are within a factor of 2 for both groups (size of $G_1$ is 44971, $\var_{h_i \in G_1}(\ell_i) = 103.94$;
size of $G_2$ is $76890$, $\var_{h_i \in G_2}(\ell_i) = 62.14$). We shall refer to $G_1$ as the small group and $G_2$ as the large group.
 Our goal is to test the following hypothesis:

\noindent{\textbf{$H_0$}}:  the small group's thread length is no greater than the larger group's thread length.

\noindent{\textbf{$H_1$}}:  the small group's thread length is larger than the larger group's thread length.

The comparison above is understood to be with respect to some central tendency measure. Figure~\ref{fig:lenthreads}b also gives the boxplot of the log of thread lengths in both groups. Using a $t$-test, we get a $t$-statistic of 40.3 and the $p$-value is $\leq 2.2\times 10^{-16}$. We also carry out a Mann-Whitney $U$-test, which also yields a $p$-value $\leq 2.2 \times 10^{-16}$. Both tests indicate that we can reject the null hypothesis with high confidence (one with respect to the mean, the other w.r.t. the median) and therefore, there is strong evidence that having more  threads created at the same time correlates with a reduction in the attention each thread receives.

\section{A generative model}\label{sec:classify}
\Znote{Replication of the results: execute Wrapper.go, gom1, gom2, gom3.}
Now that we have explored Q1, we move on to Q2. Specifically, 
this section presents a generative model for thread discussions. We start with a motivating example: evaluation of standard classifier algorithms for filtering  small-talk (recall that this is important in the initial stage of a course; see Section~\ref{sec:intro}). Specifically, it is well-known that both naive Bayes (NB) and support vector machine (SVM) classifiers are good in this type of classification problems. While the SVM classifier shows reasonable performance, the NB classifier experiences excessively high false positive rates. We may leverage the clues from this discrepancy in performance to design a generative model. This generative model both explains the discrepancy and guides us in designing topic extraction and relevance-ranking algorithms (Section~\ref{sec:ranking}). 

\subsection{Understanding the classifiers.}
\myparab{Getting the data.}
We used MTurk to label the threads.  
We randomly selected 30 threads for each course. We also chose an arbitrary subset of 40 courses and labelled the first 30 even-numbered threads. This allowed us to get more thread samples that fall into the small-talk category. 



We next split the data set into a training set and a test set: for each randomly sampled thread, with probability 0.85, we assign it as a training data point and otherwise a test data point. Since the test data set is smaller, the authors also made an extra pass on the test set to further reduce the errors in the labels. Also, the test size appears to be sufficiently large so we do not use $k$-fold cross validation here.  

\myparab{Training the classifiers.} The NB algorithm is implemented based on~\cite{Mitchell1997,Harrington2012}. 
We took two approaches to train the NB classifier: Approach 1: We train a single classifier on all of the training data. Approach 2: We train the NB classifier per course and test the classifier to the testing data in the same course.  For SVM, we use the standard Kernel. We use the open source code SVM Light~\cite{Joachims1999} software, with their default parameters.

\myparab{Results.} Figure~\ref{fig:svmvsbayes} compares the performance between the SVM and the NB classifiers. Here, it is more appropriate to measure true and false positive rates instead of precision/recall (see~\cite{Harper09} for a discussion).  Both the true positive and false positive rates are very high in NB classifiers while SVM has a substantially better false positive rate when the true positive rate is similar (the first row in Fig.~\ref{fig:svmvsbayes}), which is further reduced when we allow the SVM to have a lower true positive rate (2nd row in Fig.~\ref{fig:svmvsbayes}). We also remark that an advantage of our classification process is that we do not need to use too many features to obtain sufficiently good results.

\begin{figure}
\begin{center}
{\small
\begin{tabular}{|l|c|c|}
\hline 
 & true positive rate & false positve rate\\
\hline
SVM ($\theta = -1.015$) & 93.3\% & 17.3\% \\
\hline
SVM ($\theta = -0.995$) & 86.7 \% & 5.9\% \\
\hline
NB (aggregate) & 96.7\% & 35.9\% \\
\hline
NB (separate) & 96.7\% & 76.8\%\\
\hline
\end{tabular}
}
\end{center}
\vspace{-.4cm}
\caption{Comparing  the SVM and the NB classifier.}
\vspace{-.1cm}
\label{fig:svmvsbayes}
\end{figure}

\subsection{A unified topical model}\label{sec:topicalmodel}
We now present a generative model for the discussion threads inspired by the above experimental results. Let $C$ be the set of $n$ words appearing in the discussion threads. Our model consists of the following distributions on $C$: 

\begin{itemize}
\item \emph{The background distribution} $\calb$ that is not topical-dependent. This refers to the set of commonly used words in the English language (but not too common that would appear in the stopword list). 
\item \emph{The topical distribution for small-talk and logistics} $\calt_0$: This models keywords that are more likely to appear in small-talk or logistics discussions. 
\item \emph{The course-specific topical distribution} $\calt_i$ for the $i$-th course. This models keywords that are more likely to appear in the discussions that are relevant to the $i$-th course. 
\end{itemize}

\myparab{Sampling a thread.} A thread in the $i$-th course is sampled in the following way:

\begin{itemize}
\item With probability $p_i$ that the thread is a logistic/smalltalk thread. 
\item Otherwise the thread is a course-specific thread. 
\end{itemize}

Here $p_i$ can be different for different courses. When a thread is a logistic/smalltalk thread, the 
words are sampled i.i.d. from $\mathcal D_0(i)$, which is defined as: with probability $(1-\epsilon)$ the word 
is sampled from $\calb$ and otherwise the word is sampled from $\calt_0$. Notice that $\cald_0(i)$'s are the same for all $i$'s. Similarly, when a thread is a course-specific thread, the words in the thread are i.i.d. samples from $\mathcal D_1(i)$ defined as:
with probability $(1-\epsilon)$ the word is from $\calb$ and otherwise the word is from $\calt_i$. 

Furthermore, for exposition purposes, let us make the following assumptions (most of which can be relaxed):
\begin{itemize}
\item \textbf{Near-uniformity in $\calb$:} for any $w \in C$, $\Pr_{\cald_i(j)}(w) = \Theta(\frac 1 n)$ for 
 all $i$ and $j$. Here, we shall imagine $C$ represents the words that are outside a stopword list but cover important topics in different courses, \eg the 200-th to 2000-th most frequent words. This assumption is justified by the heavy tail distribution of English words. 
\item \textbf{Distinguishability of the topics:} Let $\supp(\mathcal D)$ be the support of
a distribution $\mathcal D$. For any $i$ and $j$,  we assume $\supp(\calt_i)$ and $\supp(\calt_j)$ 
do not overlap (but the supports of $\mathcal D_1(i)$ and $\mathcal D_1(j)$ can still overlap). Furthermore, for any $w \in \supp(\calt_i)$, $\ell \leq \frac{\Pr_{\cald_1(i)(w)}}{\Pr_{\calb}(w)} \leq u$, where $1 < \ell < u$ are two constants. 
\end{itemize} 

We have the following theorem that explains the behavior of the classifiers: 

\begin{theorem}\label{thm:topical} Consider the generative model above. There exist $s$, $\epsilon$, $p_1, ..., p_m$, and a sequence $b_1, ..., b_m$, such that if we get $b_i$ training samples for the $i$-th course, then 
\begin{enumerate}
\item With constant probability over the training set, the NB classifier will have poor performance for some courses (\ie whp the classifier errs at the negative threads) regardless of whether the classifier is trained per course or across all courses. 
\item There exists a good separation plane for SVM so that whp a discussion thread will be classified correctly. 
\end{enumerate}
\end{theorem}

Here we briefly sketch the intuition for the first part of the theorem. The full paper presents the proof. 
We leverage a quite straightforward intuition to construct the negative examples: when we train the classifier per course, there is insufficient 
training data (\eg $\approx 30$ in our case) and thus with constant probability we overestimate the conditional 
probability. On the other hand, when we train the classifier against all the courses, 
the classifier cannot address the fact that $p_i$ are different for different courses. Thus, the NB classifier will  use a wrong prior for some courses. In these courses, the classifier could perform badly. 

\section{Topic extraction and ranking.}\label{sec:ranking}
We now demonstrate how the generative model can be used to extract forum keywords and design relevance ranking algorithms. We shall start with a simple algorithm for extracting $\supp(\mathcal T_i)$, \ie the topical words for the $i$-th course. We first describe the need for this algorithm. 

\myparab{Motivation.} Knowing $\supp(\mathcal T_i)$ gives a forum user knowledge of the topics in the course. There are some other possible approaches to extract the keywords, such as using the keywords in a course syllabus. We argue that these existing approaches are not suitable solutions here. 

First, while the keywords in the course syllabus appropriately summarize the course lectures, they do not always summarize \emph{the course discussions}. For instance, in a freshman English writing course, the instructors may focus on ``high level'' issues of writing, such as organizing essays or delivering effective arguments. The discussions in the forum may focus on ``lower level'' issues or other relevant questions, such as the usage of punctuation.  Second, when the same course is offered multiple times, we expect the topics of the forum discussion would be different and thus $\mathcal T_i$ also changes. 

\myparab{The topic extraction algorithm.} Let $i$ reference a particular course. Our algorithm uses the following two parts of training data:
\begin{itemize}
\item \textbf{The background training data} consists of forum discussions of $k$ courses. 
\item \textbf{The course-specific training data} consists of forum discussions in the $i$-th course in the first few days (approximately 10 days). 
\end{itemize}

Let $n$ be the total number of words we see in the background training data set, and let $\hcald_n$ be the empirical unigram distribution associated with both the background and course-specific training data. Let  $\hcale$ be the empirical distribution associated with the course-specific training data. Let $W = \{w_1, ..., w_{\ell}\}$ be the support of $\hcale$, and $w$ be an arbitrary word in $W$.  
Let $p_{\hcald}(w)$ be the probability mass of $w$ under the distribution $\hcald_n$; let $p_{\hcale}(w)$ be the probability mass of $w$ under $\hcale$. We define the ``surprise weight'' of $w$ as 
$$\gamma(w) = \frac{p_{\hcale}(w)n}{\sqrt{p_{\hcald}(w)n}} = \frac{p_{\hcale}(w)\sqrt n}{\sqrt{p_{\hcald}(w)}}$$

We next rank the words in $W$ according to the $\gamma(\cdot)$ function, \eg the first word $w$ has the highest $\gamma(w)$. Our summary of keywords is the top-$k$ (ordered list of) words in the ranking.  We have the following Corollary. 

\begin{corollary} Under our generative model presented and assuming $p_i = \Theta(1)$ and  $k = |\supp(\mathcal T_i)|$ is known, our topical extraction algorithm will successfully identify $\supp(\calt_i)$  when the training data is sufficiently large. 
\end{corollary}

\myparab{Experiments.} We now evaluate the efficacy of our topical algorithm. Here, the main discovery is that we need only approximately 10 days of course-specific training data for the algorithm to accurately identify $\mathcal T_i$. 

We tested the algorithm on 10 random ``large'' courses. Specifically, we first identify the set of courses that have larger discussion volume: for each course, we count the number of days, in which 15 or more new threads are created. Call this number $d_i$ for the $i$-th course. We then find all courses with $d_i \geq 25$. There are a total number of 24 such courses. Finally we randomly choose 10 courses for testing.  We then choose 50 courses among the remaining courses as background training courses. 

Figure~\ref{fig:keywordresult} shows the results of our topical algorithm for four courses (the rest of the  results are in the full technical report). We can see that the topical model is quite effective in identifying the keywords across all test courses. We also remark that the terms with the highest idf score are mostly meaningless words. 

\begin{figure*}
\begin{center}
\includegraphics[scale=0.2]{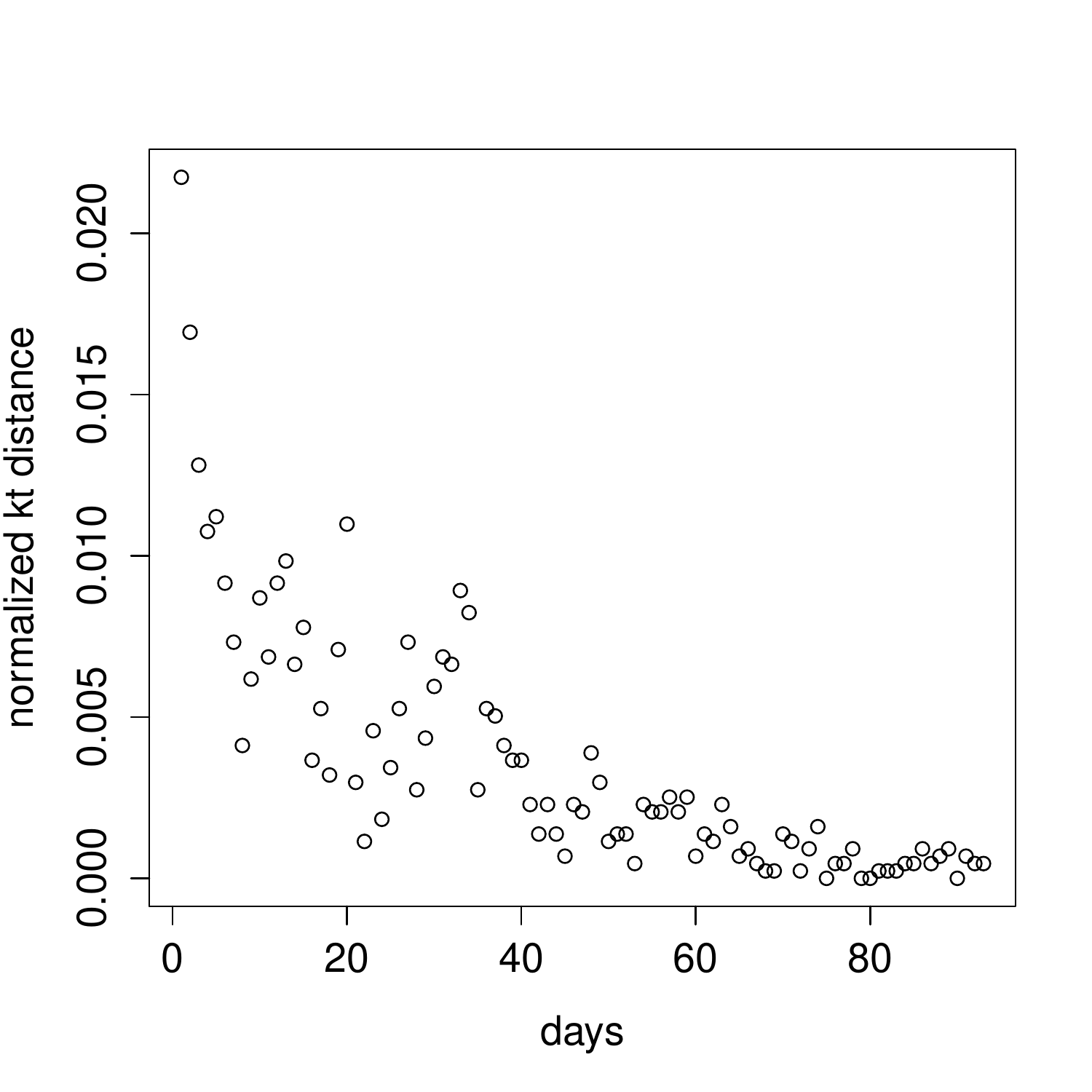}
\includegraphics[scale=0.2]{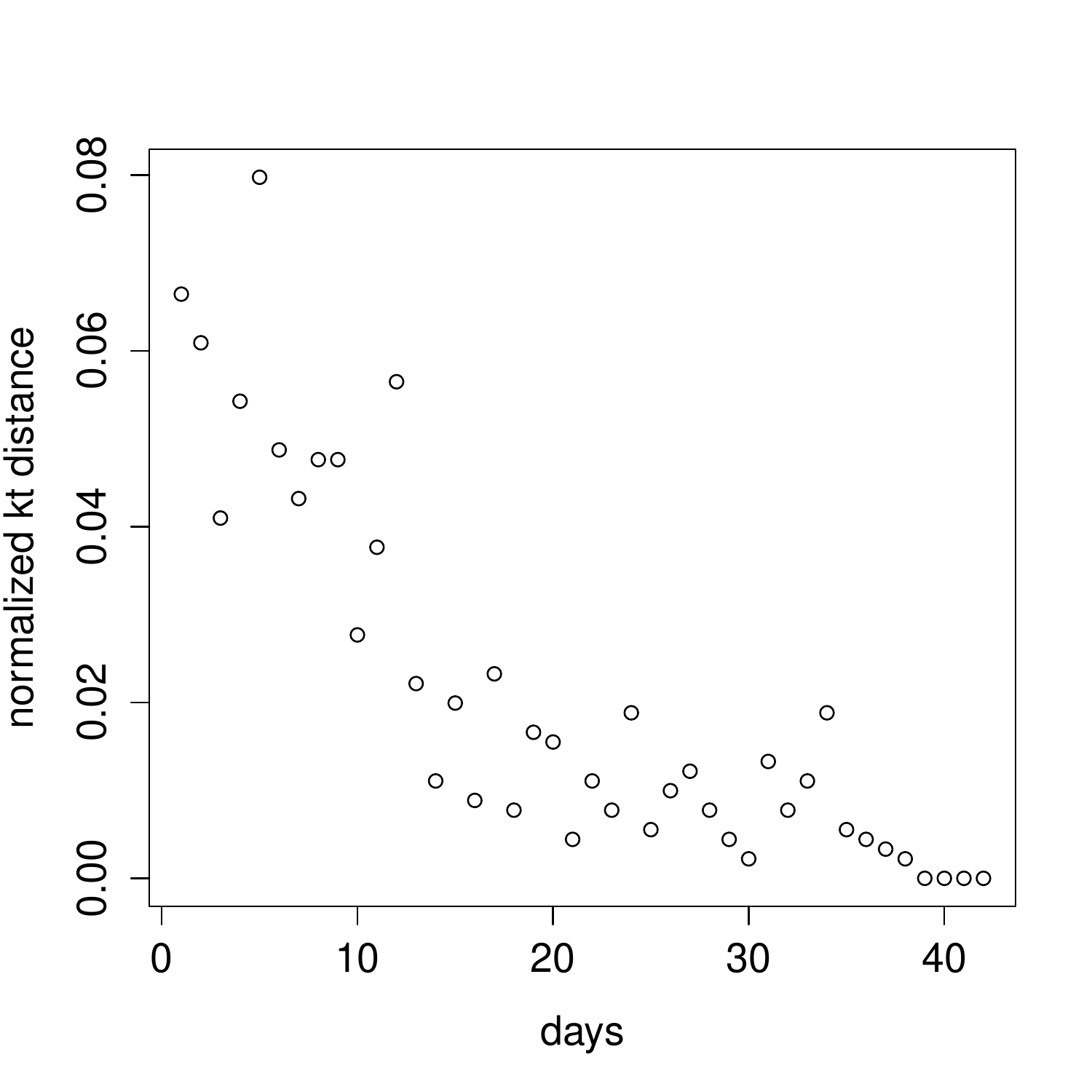}
\includegraphics[scale=0.2]{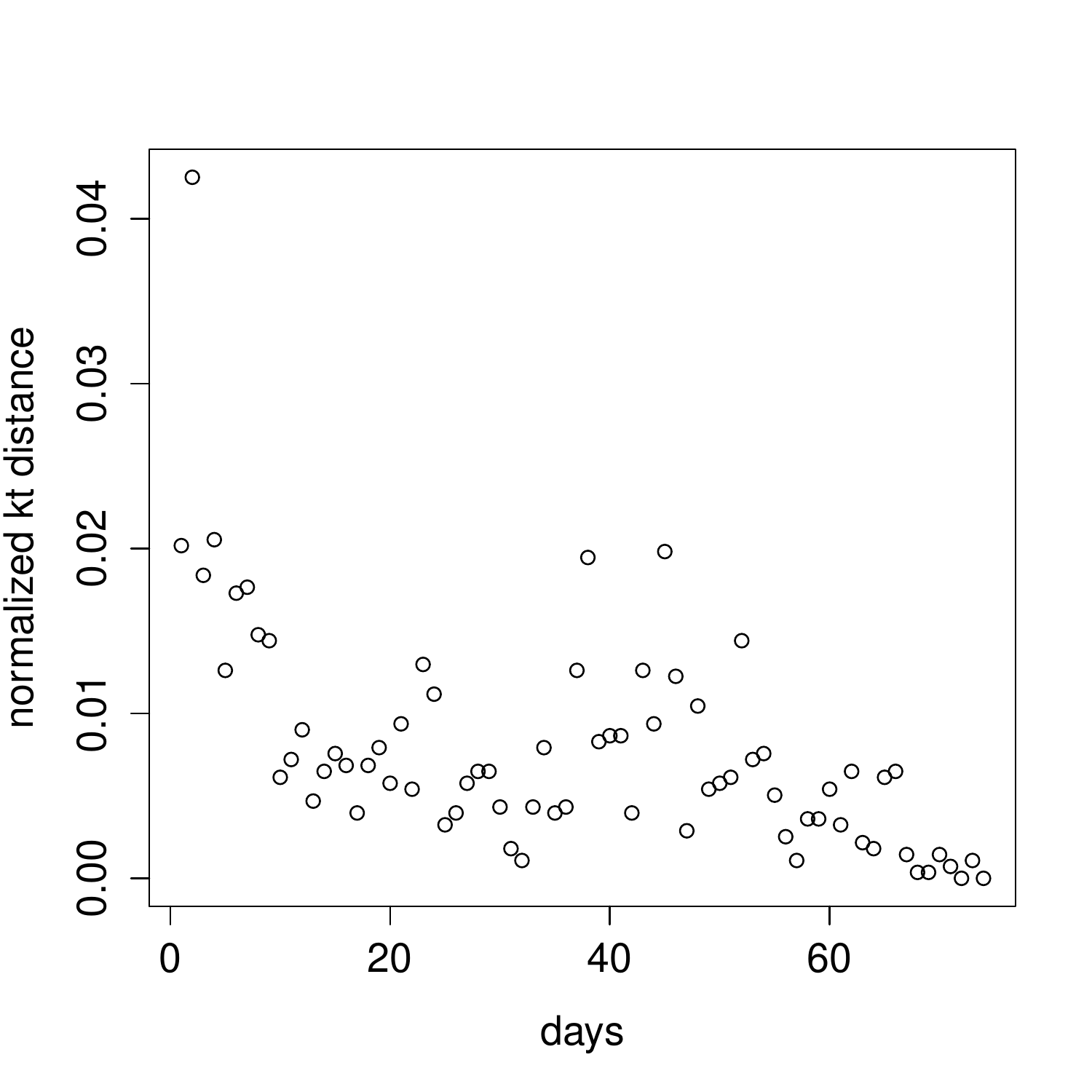}
\includegraphics[scale=0.2]{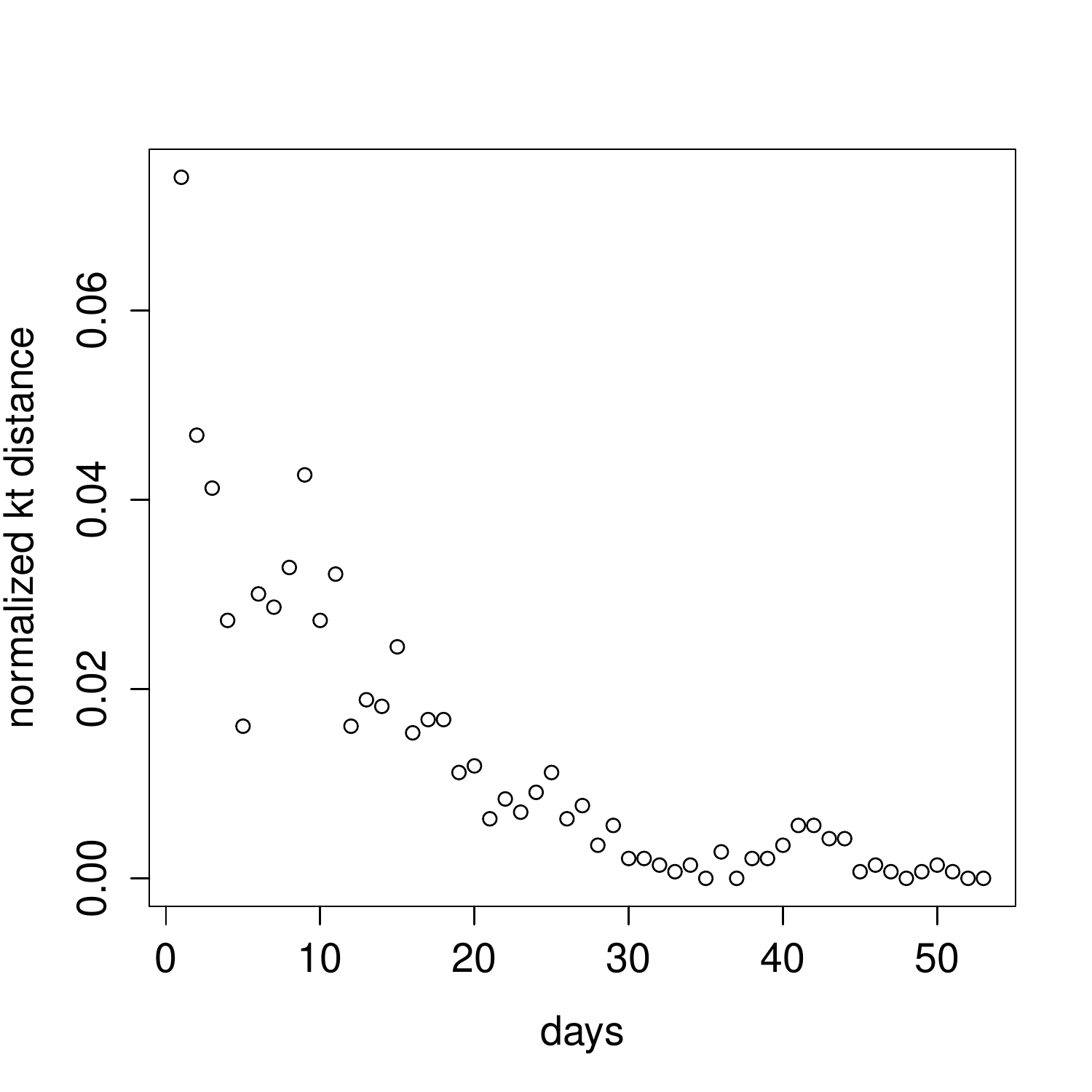}
\end{center}
\vspace{-.3cm}
\caption{Our algorithm's convergence rate for {\small ``Machine Learning'', ``Health for All Through Primary Health Care'', ``Pattern-Oriented Software Architectures'', ``Latin American Culture''} (from left to right).}
\label{fig:topicconverge}
\end{figure*}

We would also like to understand the convergence rate of the algorithm. This corresponds with the time window for ``warming up'' the algorithm. The sets of top-50 keywords quickly stabilize for different courses (see full paper). 
Figure~\ref{fig:topicconverge} presents the normalized Kendall tau distance for ranks between two consecutive days for the same set of courses (the normalized Kendall tau distance is defined by the Kendall tau distance divided by the total number of distinct word pairs). 
We can see that the Kendal tau distance also quickly converges to below 2\% after approximately 10 days. 
\begin{figure*}
\begin{center}
{\footnotesize
\begin{tabular}{|p{3.5cm}|p{13cm}|}
\hline 
  Course name & Keywords \\
\hline
{\small Machine Learning} & {\small theta	octave	regression	gradient	descent	matrix	ex1	alpha	machine	vector	function	linear	computecost	gnuplot	data	}
	\\
\hline 
{\small Health for All Through Primary Health Care} & {\small health	phc	care	primary	healthcare	community	ata	alma	perry	henry	medical	clinics	services	public	communities}	\\
\hline
{\small Pattern-Oriented Software Architectures }  & {\small doug	vanderbilt	patterns	dre	schmidt	posa	concurrent	concurrency	gof	middleware	corba	frameworks	pattern	software	singleton}\\
\hline
{\small Latin American Culture} &{\small latin	america	culture	indigenous	american	cultures	democracy	spanish	imposed	political	cultural	traditions	countries	democratic	mexico}	\\
\hline
\end{tabular}
}
\end{center}
\vspace{-.3cm}
\caption{Results of our algorithm for extracting the keywords (top 15 keywords.)}
\label{fig:keywordresult}
\end{figure*}

\myparab{Ranking.} We next leverage our topic-extraction algorithm to design a relevance-ranking algorithm. 

To begin, let us walk through a hypothetical scenario to set up the notation. Alice enrolled in a machine learning course and she used the forum on the 12th day. Then on the 15-th day, Alice logs in again. Here, our goal is  to rank the new threads that are created between the 12th and the 15th day. We shall refer to the first 12 days as the \emph{warming period}. The 12th - 15th days are the \emph{query period}. 
Finally, these  15 days are referred to as the \emph{window of interest} for the ranking algorithm. 

We next describe our algorithm. Roughly speaking, the algorithm puts more weight
on the threads that contain more keywords from $\supp(\mathcal T_i)$. More specifically, the algorithm consists of two steps: \emph{Step 1. Assignment of weights.}  Let $w$ be an arbitrary keyword that appears in the thread. Let $r(w)$ be the rank of $w$ in the output of our topical algorithm. In case $w$ is not  in the top-50 most frequent keyword set, then $r(w) = \infty$. The weight of $w$, $\eta(w)$, is $\alpha^{r(w)}$, where $\alpha$ is a parameter between $0$ and $1$ ($\alpha = 0.96$ in our experiments).  \emph{Step 2. Assigment of scores.} The score of a thread is simply the sum of the weight of each word (with repetition) that  appeared in the thread. 

While our algorithm is technically simple and shares similar spirit with  tf-idf based algorithms~\cite{Manning2008}, we emphasize the primary goal here is a proof-of-concept for the efficacy of our generative model. 

We next compare our algorithm with two natural baseline algorithms, a term-based one (tf-idf) and a ``random walk'' based one. 

\mypara{The tf-idf based algorithm.} In the tf-idf based algorithm, we treat each thread as a document and the ranking simply sorts the threads in descending order by the tf-idf scores. 

\mypara{Kleinberg's HITS type algorithm.} One possible random walk-based algorithm is a variation of the HITS algorithm~\cite{Kleinberg1999}, which is good at finding the ``importance'' of the nodes in a network and works as follows. First, we construct a bipartite graph: each node in the LHS represents a user and each node in the RHS represents a thread. A user node is connected with a thread node if and only if the user participates  in that thread. Then we interpret the thread nodes as ``authorities'' and the user nodes as ``hubs'' and apply  the standard HITS-type updates (see the full technical report). Finally, we give a rank on the threads by their authority scores. 

\myparab{Comparisons.} Before we present our experimental results, we  highlight the differences between our algorithm and the baseline ones. 

\noindent{\emph{Comparing with tf-idf.}} (1) Our algorithm considers all the threads in a course while the tf-idf technique focuses on a single thread. Thus, the tf-idf technique could possibly pick up threads that contain low-frequency words which are irrelevant to the course, including discussion in non-English languages, or students in a medical course 
soliciting a doctor's opinion on his/her sick acquaintance or other similar scenarios. (2) The tf-idf technique is incapable of 
finding important keywords for the forums because the terms with the highest idf scores are often those that appear exactly once, most of which do not have much meaning. So it has less ``interpretability''. 

\noindent{\emph{Comparing with HITS.}} While HITS can find the ``popular'' threads, our goal here is to find the relevant ones. We do not expect graph-based algorithms work well without looking into discussions' content. 

\myparab{Evaluation.} We now validate the above claims by experiments. Specifically, our goal is to validate (1) the tf-idf based algorithm would  have a higher probability mis-weighting the non-relevant threads, such as non-English discussions. (2) The HITS-based algorithm will also give more irrelevant threads because it ranks  popularity instead of  relevance. 

\noindent{\emph{Testing and training courses.}} We test on the same 10 ``large courses'' as we did for the keyword extraction algorithm. 

\noindent{\emph{Choosing the windows of interest.}} We randomly choose 5 days from the 10th to the 30th day to form a set $D$. And then we also add the 10th day to the set. For each $d \in D$, we create a test case with $d$ days of warming period and 2 days of query period. 

\noindent{\emph{Direct comparison.}}  It is a major open problem to directly assess the quality of multiple ranks in an efficient manner in computational social choices (see ~\cite{AzariSoufianiuai13} and references therein). 
Thus, we focus on understanding the \emph{differences} between our algorithm and the baseline ones by comparing the number of \emph{irrelevant} threads that are recommended by the algorithms.
 
To do so, for each course and each window of interest, we pull out the set of the first 15 threads recommended by our algorithm $S$ and the set of the first 15 threads recommended by the baseline algorithm $S_b$. Then we find the differences $D_1 = S - S_b$ and $D_2 = S_b - S$. Next, we use MTurk to label whether the threads in $D_1$ and $D_2$ are relevant or not. This will give us difference in relevance counts between algorithms.

\Znote{Files are in prediction/comp1outputb.txt and prediction/comp2outputb.txt}

\myparab{Result 1. Comparing with tf-idf based algorithm.}  Here, $|D_1| = |D_2| = 253$ for the 10 courses and 6 days we examined.  64 were labeled as irrelevant in $D_1$ and 104 were irrelevant in $D_2$. 

\begin{figure}
\begin{center}
\subfigure[Comparing wt tf-idf]{
\hspace{-.5cm}
\includegraphics[scale=0.33]{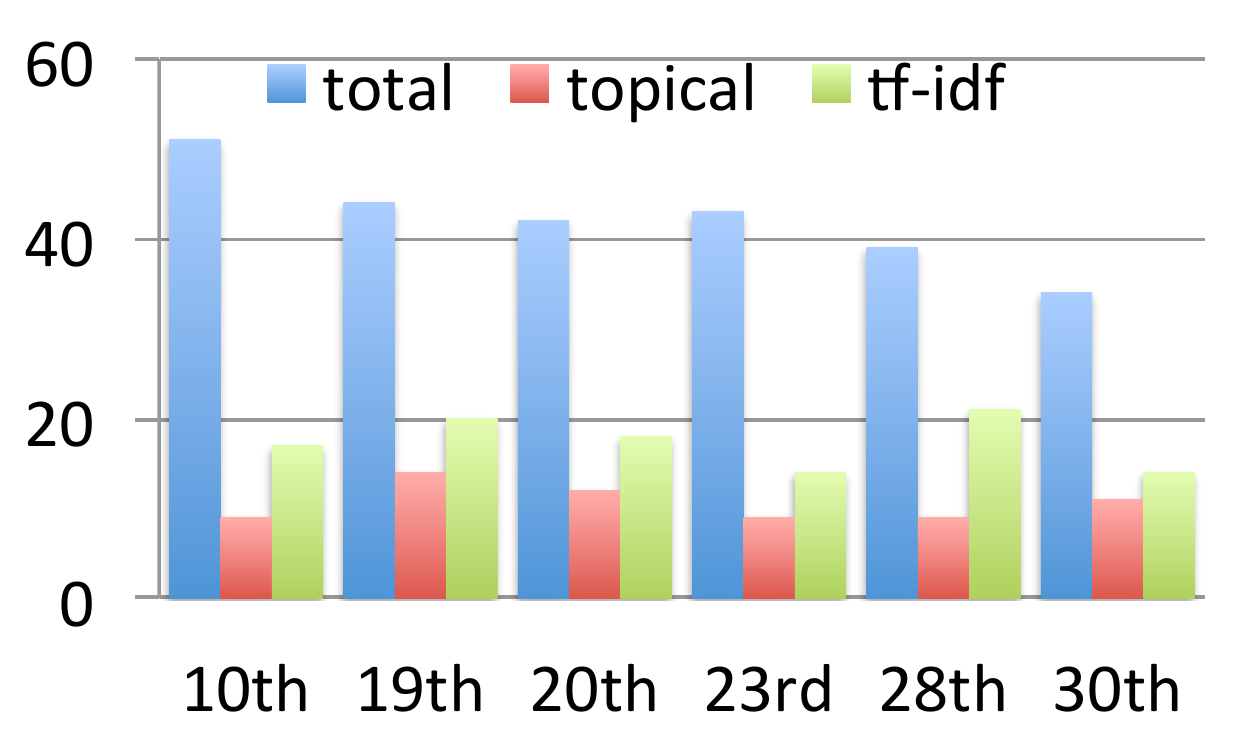}}
\subfigure[Comparing wt HITS]{
\includegraphics[scale=0.33]{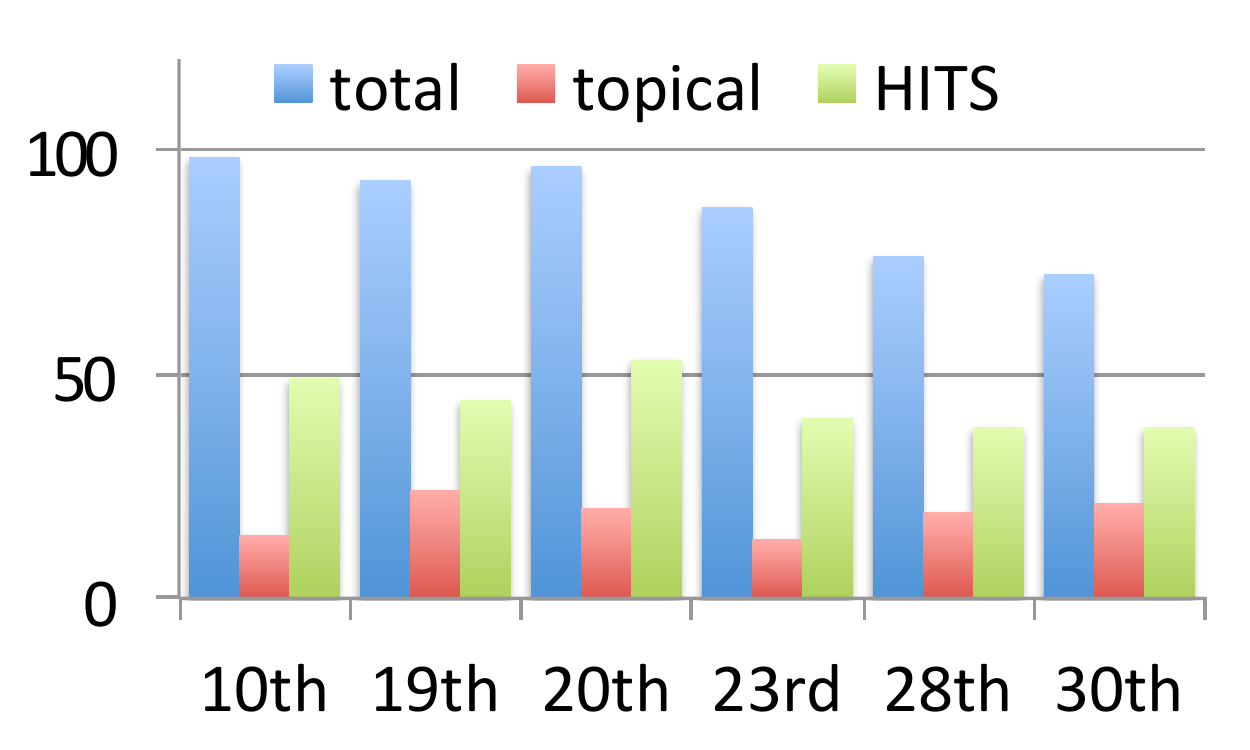}}
\end{center}
\vspace{-.4cm}
\caption{Evaluating our relevance-ranking algorithms.}
\label{fig:comp1}
\end{figure}

Figure~\ref{fig:comp1}a shows the breakdown of the misclassified threads by days (full report also presents  the plot by courses). 
The blue bar is the total size of the differences in each day for all courses. The red bar represents the number of irrelevant threads from our algorithm and the green bar represents the number of irrelevant threads from the tf-idf one. We can see that our algorithm is consistently better. While the improvement is not significant, this validates our intuition discussed above. 

\myparab{Result 2. Comparing with HITS.} 
In this case $|D_1| = |D_2| = 522$. 111 threads were irrelevant in $D_1$ and 262 threads were irrelevant in $D_2$. Figure~\ref{fig:comp1}b shows further breakdown by days (the full paper has the plot by courses). Our algorithm is again consistently better and the difference is more substantial than result 1. This validates that the HITS-type algorithm is less effective in finding relevant threads. 

\section{Conclusion}
The larger goal behind our two main research questions is to improve the quality of learning via the online discussion forums, namely by (1) sustaining forum activities and (2) enhancing the
personalized learning experience. This paper makes a step
towards achieving these end-goals by relying on an extensive empirical
dataset that allows us to understand current user behavior as well as factors that could potentially change the current user behavior. We showed, for example, that the teaching staff's
active participation in the discussion increases the discussion
volume but does not slow down the decline in participation.
We also presented two proof-of-concept algorithms for keyword extraction and relevance-ranking to remedy the information overload problem, both of which are demonstrated
to be effective. Devising effective strategies to reduce the
decline of participation in the online discussion forums is
the main open problem to be addressed next.
\vspace{0.2cm}

{
\bibliographystyle{abbrv}
\bibliography{MOOClearning}
}
%
%
\end{document}